\documentclass[traditabstract]{aa} 
\usepackage{graphicx}
\usepackage{psfrag}
\usepackage{amsmath}
\usepackage{amssymb}
\usepackage{color,xcolor}
\usepackage{hyperref}
\usepackage{caption,natbib}
\hypersetup{
    colorlinks=true,
    citecolor=blue,
    filecolor=black,
    linkcolor=blue,
    urlcolor=magenta,
    linktocpage=true,
    breaklinks = true
}

\usepackage{enumitem}
\usepackage{subfigure,float}
\usepackage{multirow}
\usepackage{makecell}
\usepackage[normalem]{ulem}

\usepackage{array,booktabs}
\usepackage{siunitx}
\DeclareUnicodeCharacter{2212}{\textendash}

\newcommand{\beq}{\begin{equation}}
\newcommand{\eeq}{\end{equation}}

\newcommand{\ber}{\begin{eqnarray}}
\newcommand{\eer}{\end{eqnarray}}

\newcommand{\ba}{\begin{align}}
\newcommand{\ea}{\end{align}}

\def \mut {\mu_{\rm true}}
\def \dtt {\Delta t_{\rm true}}

\def \dt{\Delta t}

\def \mutil {\tilde{\mu}}
\def \ttil {\tilde{t}}

\begin{document}

   \title{Detecting unresolved lensed SNe Ia in LSST using blended light curves}

   \author{Satadru Bag\inst{1,2}\thanks{satadru.bag@tum.de}, Simon Huber\inst{1,2}, Sherry H.~Suyu\inst{1,2,3}, Nikki Arendse\inst{4}, Irham Taufik Andika\inst{1,2},
Raoul Ca{\~n}ameras\inst{2,1,5}, Alex Kim\inst{6}, Eric Linder\inst{6,7}, Kushal Lodha\inst{8,9}, Alejandra Melo\inst{1,2},
Anupreeta More\inst{10,11}, Stefan Schuldt\inst{12,13} \and Arman Shafieloo$^{8,9}$
          }

   \institute{Technical University of Munich, TUM School of Natural Sciences, Physics Department,  James-Franck-Stra{\ss}e 1, 85748 Garching, Germany
         \and 
           Max-Planck-Institut f{\"u}r Astrophysik, Karl-Schwarzschild Stra{\ss}e 1, 85748 Garching, Germany
           \and
           Academia Sinica Institute of Astronomy and Astrophysics (ASIAA), 11F of ASMAB, No. 1, Section 4, Roosevelt Road, Taipei 10617, Taiwan
           \and
          Oskar Klein Centre, Department of Physics, Stockholm University, SE-106 91 Stockholm, Sweden
          \and
          Aix Marseille Univ, CNRS, CNES, LAM, Marseille, France
          \and
          Lawrence Berkeley National Laboratory, 1 Cyclotron Road, Berkeley, CA 94720, USA
          \and
          Berkeley Center for Cosmological Physics, University of California, Berkeley, CA 94720, USA
          \and
          Korea Astronomy and Space Science Institute (KASI), 776 Daedeok-daero, Yuseong-gu, Daejeon 34055, Korea
          \and
          KASI Campus, University of Science and Technology, 217 Gajeong-ro, Yuseong-gu, Daejeon 34113, Korea
          \and
          Inter-University Centre for Astronomy and Astrophysics, Post Bag 4, Ganeshkhind, Pune 411007, India
          \and
          Kavli Institute for the Physics and Mathematics of the Universe (IPMU), 5-1-5 Kashiwanoha, Kashiwa-shi, Chiba 277-8583, Japan
          \and
          Dipartimento di Fisica, Universit\`a  degli Studi di Milano, via Celoria 16, I-20133 Milano, Italy
          \and
          INAF - IASF Milano, via A. Corti 12, I-20133 Milano, Italy
}


 \abstract
{
Strongly gravitationally lensed supernovae (LSNe) are promising probes for providing absolute distance measurements using gravitational-lens time delays. Spatially unresolved LSNe offer an opportunity to enhance the sample size for precision cosmology. We predict that there will be approximately three times as many unresolved as resolved LSNe Ia in the Legacy Survey of Space and Time (LSST) by the {\it Rubin} Observatory.  In this article, we explore the feasibility of detecting unresolved LSNe Ia from a pool of preclassified SNe Ia light curves using the shape of the blended light curves with deep-learning techniques. We find that $\sim 30\%$ unresolved LSNe Ia can be detected with a simple 1D convolutional neural network (CNN) using well-sampled $rizy$-band light curves (with a false-positive rate of $\sim 3\%$).
Even when the light curve is well observed in only a single band among $r$, $i$, and $z$, detection is still possible with false-positive rates ranging from $\sim 4$ to $7\%$ depending on the band. Furthermore, we demonstrate that these unresolved cases can be detected at an early stage using light curves up to $\sim20$ days from the first observation with well-controlled false-positive rates, providing ample opportunity to trigger follow-up observations.
Additionally, we demonstrate the feasibility of time-delay estimations using solely LSST-like data of unresolved light curves, particularly for doubles, when excluding systems with low time delays and magnification ratios. However, the abundance of such systems among those unresolved in LSST poses a significant challenge. This approach holds potential utility for upcoming wide-field surveys, and overall results could significantly improve with enhanced cadence and depth in the future surveys.
}
   {}
   {}
   {}
   {}


   \titlerunning{Unresolved lensed SNe Ia in LSST}
   \authorrunning{Bag et al}

   \maketitle

\section{Introduction}

Gravitationally lensed transients, such as quasars (QSOs) and supernovae (SNe), have emerged as novel powerful cosmic probes capable of delivering cosmological information independently of other probes (see, e.g.,\citet{Treu:2022aqp} for a review).  For instance, one can estimate the Hubble constant ($H_0$) using lensed transients without relying on additional datasets \citep{Refsdal1964_2,Refsdal1964_1,Saha,Oguri_2007, Bonvin2017,Wong:2019kwg,2020A&A...643A.165B,Birrer,Kelly:2023mgv,Grillo:2024rhi,Pascale:2024qjr}; importantly, the value of $H_0$ remains quite insensitive to the chosen cosmological model. This capability can address the significant tension observed in $H_0$ calculations between early Universe probes \citep[inverse distance ladder,][]{Planck:2018vyg} and late Universe probes \citep[distance ladder,][]{Riess2022}. With physically motivated lens mass models, the H0LiCOW team has further demonstrated that $H_0$ can be tightly constrained by combining constraints from multiple strongly lensed QSOs \citep{Wong:2019kwg}, thus underscoring the importance of having a large sample size.

Among variable sources, QSOs and SNe are currently the most prominent types, each possessing distinct advantages. For instance, lensed QSOs are more abundant and, as opposed to SNe, do not fade away,  making them more reliable for time-delay cosmography at present. On the other hand, lensed SNe (LSNe) are poised to take the forefront in the coming decade due to a better understanding of their light curves, feasibility of improved stellar kinematic and SN host-galaxy measurements after the SN fades away \citep{Ding2021, Suyu_2024}. Additionally, type Ia SNe serve as standard candles, offering direct luminosity distances to the sources, provided that magnification effects of microlensing (lensing by stars in the foreground lens galaxy) are mitigated \citep{Foxley-Marrable2018,Weisenbach2021}. However, a significant challenge stems from the rarity of LSNe, with only a handful of confirmed detections to date \citep{Kelly:2014mwa,Goobar:2016uuf,More2017,2021NatAs.tmp..164R,chen2022,Goobar2023,Frye2023,sn_encore}. Furthermore, many LSNe will remain unresolved in the wide-field surveys due to their angular resolution being limited by seeing constraints \citep{Quimby2014,Goldstein2018}. Detecting these unresolved LSNe could notably enhance the sample size, thereby helping us to achieve precision cosmology. Therefore, the importance of detecting  unresolved LSNe in time-domain wide-field surveys, such as the Legacy Survey of Space and Time (LSST) by the {\it Vera C. Rubin} Observatory, cannot be overstated.

Various techniques, such as the magnification method \citep{Goldstein_2017,Goobar:2016uuf, Wojtak2019}, colour--magnitude diagram \citep{Quimby2014,Nikki_2023}, and shifting centroid \citep{Ramanah2022}, can detect unresolved LSNe from photometry. As the observed light curve of an unresolved LSN is the sum of the individual image light curves, the former contains distinct features in its shape and can also be used to discern the lensed systems from the unlensed ones. This constitutes the central focus of our work. \footnote{Recently, similar efforts have been undertaken to detect unresolved lensed QSOs from light-curve observations \citep{Shu2020,Bag2021qso,Bag_qso_2023}. However, it is important to note that the techniques employed are entirely different due to the distinct nature of QSO light curves.} We developed several methods based on forward modelling for this purpose \citep{Bag:2020pbg,misha2021}. In the present article, inspired by \citet{misha2022}, we employ a simple 1D convolutional neural network (CNN) that leverages the shape of the light curves to detect lensed cases. We apply the CNNs to light-curve observations consistent with those achievable with LSST by {\it Vera C. Rubin} Observatory. LSST, which is deeper than the ongoing Zwicky Transient Facility (ZTF) survey \citep{ztf}, is expected to observe approximately 20,000 square degrees over 10 years \citep{lsst1,lsst2,LSST_2018}, thereby notably increasing the potential for discovering new unresolved LSNe.

Our primary emphasis is on Type Ia supernovae (SNe Ia), given their pivotal role in cosmology as standardisable candles. We simulate mock LSST light curves using the \texttt{baseline v3.2} observing strategy. The sample comprises joint light curves of unresolved LSNe Ia as positive examples and unlensed (i.e. intrinsic) SNe Ia light curves as negative examples. We employ two types of  CNN: one for detection and the other intended to estimate time delays directly from the joint light curves in cases when a follow-up observation is not possible or conclusive.
 It is worth noting that this approach relies on the precise photometric classification of SNe Ia from other transients, for example using the algorithms proposed by \citet{Lochner2016,Alves22}, prior to applying the approach.

The article is organised as follows. In Section 2, we define the unresolved cases in LSST and compare their forecasted numbers and properties against the resolved ones. Next, in Section 3, we explain the process of simulating mock LSST light curves. Section 4 presents a brief overview of the method and data processing. Our main results are explained in Section 5. Finally, in Section 6, we conclude and discuss further scopes in this direction.

\section{Unresolved SNe Ia in LSST}
\label{sec:1}

\subsection{Unresolved light curves}
\label{sec:unres_lc}

The blended joint light curve of an unresolved LSN with $N$ images is the sum of the individual image fluxes (neglecting microlensing for the moment),
\beq \label{eq:unres_lc1}
F(t)=\sum_{i \in \rm images}^N |\mutil_i| \mathcal{F}(t-\ttil_i)\;,
\eeq
where $\mutil_i$ and $\ttil_i$ are the magnification (which can be positive or negative depending on the parity of an image) and time delay of the $i$th image. Note that the light curve of all the images can be described by the same function $\mathcal{F}(t)$ (the unlensed light curve; here we ignore microlensing and differential dust extinction for simplicity) but with different magnifications and time delays. It is useful to recast the above expression in terms of the image that arrives the earliest, which we refer to as the `first' image,
\beq \label{eq:unres_lc}
F(t)=\sum_i^N \mu_i f(t-\dt_i)\;,
\eeq
where $f(t)=|\mutil_1| \mathcal{F}(t-\ttil_1)$ is the light curve of the first image and
\beq
\mu_i \equiv |\mutil_i/\mutil_1| ~~{\rm and}~~ \dt_i \equiv \ttil_i-\ttil_1
\eeq
represent the magnification ratio and time delay of the $i$-th image with respect to the first image, respectively. By definition $\dt_i,~ \mu_i \geq 0$ for all the images and naturally, $\mu_1=1$ and $\dt_1=0$.

In this work, we only consider the light-curve information and aim to detect whether or not the system is lensed, that is, if the observed light curve stems from the blending of multiple images or corresponds to an unlensed SN. It is important to note that the task becomes challenging when dealing with extreme values of $\mu$ or small values of $\dt$. To understand this issue, let us first consider the double systems for simplicity. When the magnification ratio deviates significantly from unity (with $\dt$ not being too high), one image dominates over the other; here, $\mu_2 \ll 1$ ($\mu_2 \gg 1$) implies that the second (first) image has negligible brightness compared to the first (second) image. On the other hand, if $\dt_2 \to 0$, the two images coincide with each other. Therefore, in both of these conditions, it is mathematically impossible to detect the lensing nature in the unresolved light curve. This can be easily generalised for quads; if two images have extreme magnification ratios or very small time delay, then they cannot be discerned as two images from the unresolved light curve.
In simpler terms, to detect lensed systems, it is essential to have at least one pair of images (which is the only pair in the case of a double) with $\mu \sim \mathcal{O}(1)$ and $\Delta t >$ a few days. Furthermore, to differentiate quads from doubles, multiple such pairs of images are required.

\subsection{OM10 simulation for LSST}
\label{sec:om10_sim}

\begin{figure}
 \centering
\includegraphics[width=0.5\textwidth]{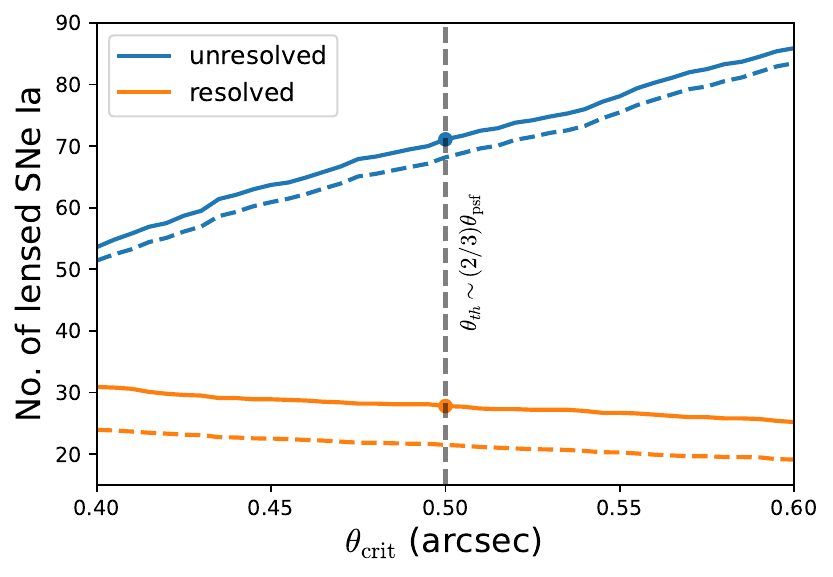}
\caption{Number of unresolved LSNe Ia (blue) is compared with the number of resolved cases (orange) for one cumulative year of LSST observation at different critical angular resolution ($\theta_{\rm crit}$) above which the images can be resolved. The solid curves are calculated assuming the macro-magnifications only, whereas the dashed curves consider the microlensing effects. LSST can resolve images up to $\theta_{\rm th} \approx 0.5 ~\text{arcsec} \sim 2/3 \theta_{\rm psf}$. For this threshold, we obtain approximately $71.1$ unresolved (blue dot) and $27.8$ resolved (orange dot) LSNe Ia based solely on macro-magnification calculation, indicating roughly $2.6$ times more unresolved systems. The ratio increases to approximately $3.2$ when we account for microlensing, as evident from the dashed curves. We emphasize that these numbers correspond to one effective year of LSST observation over $20,000$ deg$^2$ in $i$-band. LSST will observe a sky location cumulatively  for $\sim 4$ years on average during its ten-year run. Therefore, it is essential to scale these figures by a factor of $\sim 4$ to get the realistic predictions.}
\label{fig:Nstat}
\end{figure}

In this work, we concentrate on samples of type Ia SNe due to their significant cosmological utility, well-understood light curves, and ease of simulation.
Using the Oguri and Marshall 2010 simulations (OM10 hereafter, \cite{om10}) with minor updates \citep{Oguri2018}, we forecast the number of lensed type Ia supernovae (LSNe Ia) that will remain unresolved in LSST. A key observation from Equation \eqref{eq:unres_lc1} is that (lensed but) unresolved sources will appear brighter than its individual images (if they were resolved) because light from different images can be accumulated for the unresolved cases. This in turn suggests that unresolved lensed sources could be visible at greater luminosity distances (or redshifts). Our primary focus is on the LSST $i$-band because LSNe Ia are at such redshifts that they are usually observed with the highest signal-to-noise ratio (S/N) in this filter \citep{om10,holismokes7}. Additionally, the LSST $i$-band presumably offers a more regular cadence, less affected by the moon phase, and typically provides better seeing conditions compared to shorter wavelengths. In the simulation, we allow all image separations and source brightness up to $30$ mag in $i$-band to consider all possible unresolved LSNe Ia that could be observed in LSST.

We define the unresolved LSNe, detectable in LSST, with the following criteria: (i) image separation has to be smaller than a critical value above which one can resolve the images: $\theta < \theta_{\rm crit}$ and (ii) peak brightness (of the joint light curve) in $i$-band has to be brighter than {\bf a} critical magnitude, $m_{\rm crit}$. The 5$\sigma$ depth ($m_5$) of LSST $i$-band is $23.3$. However, we need the SN to be brighter than this limiting magnitude in order to have light-curve
observations of sufficiently good quality to allow time-delay measurements with adequate accuracy. Therefore, we set the critical magnitude to be $0.7$ magnitudes brighter (i.e. lower) than the limiting magnitude \citep{om10,Dimitriadis2019}, i.e. $m_{\rm crit}=22.6$.

On the other hand, we define the resolved systems as those that have at least one pair of images with angular separation larger than the critical value, $\theta > \theta_{\rm crit}$ , with both images being brighter than $m_{\rm crit}=22.6$ \footnote{In order to make a fair comparison, it is unimportant whether it is a double or a quad at this moment; this can be established during follow up.}. Figure \ref{fig:Nstat} illustrates how the number of resolved (orange) and unresolved (blue) LSNe Ia varies with the critical angular separation ($\theta_{\rm crit}$) for one cumulative year of LSST observations. Therefore, this figure presents the theoretical prediction of the number of resolved and unresolved LSNe Ia expected to occur within a year (irrespective of the observing strategy) across the LSST footprint of $20,000$ deg$^2$. Not all of these events will be observed due to seasonal gaps. Over its 10 year operation, LSST will, on average, cumulatively monitor each sky location for approximately 4 years. Therefore, for a realistic prediction, it is necessary to scale all the curves by a factor of approximately $\sim 4$.

The solid curves in Figure \ref{fig:Nstat}  assume only the macro-magnifications, whereas the dashed curves are obtained by considering (rough) microlensing effects in the light curves \footnote{OM10 simulation for 10 years produces too many LSNe Ia for it to be computationally feasible to calculate microlensing for all. Therefore, we take a representative subset of systems to generate the microlensing magnification maps, and for each system in the OM10, we pick the most similar system in this representative subset to approximate the microlensing effect.  For each system, we overpopulate it by a factor of 100 and take 100 random positions on its corresponding microlensing magnification maps to scale the fluxes of the multiple images in the system.  After repeating this procedure for all systems in OM10, we renormalize by the factor of 100.}. For this range of $\theta_{\rm crit}$, we always obtain a few times more unresolved systems than resolved ones. Note that the sum of the number of unresolved and resolved cases is not constant, as the latter must meet stricter criteria. Specifically, for resolved cases, at least one pair of lensed images must be brighter than $m_{\rm crit}$, while for unresolved cases, it is the combined flux that has to be brighter. In other words, when we lower $\theta_{\rm crit}$ by a certain amount, some of the previously unresolved systems will now have $\theta>\theta_{\rm crit}$ and the number of unresolved cases will decrease; however, not all of them will be considered as resolved, resulting in smaller growth in the resolved case number.

The angular resolution of wide field surveys, such as LSST, is limited by seeing. For LSST, \citet{om10} adopt an angular resolution of $\theta_{\rm th}=0.5$ arcsec ---which is roughly two-thirds of the median $\theta_{\rm psf} \approx 0.7$ arcsec--- as the minimum separation that could be resolved in LSST. Following this, we also set $\theta_{\rm crit}=\theta_{\rm th}=0.5$ arcsec for our estimations of resolved and unresolved cases. The final definitions of unresolved and resolved LSNe Ia are as follows:\begin{itemize}
    \item {Resolved cases} are those for which we have at least one pair of images with (i) angular separation of $\theta > 0.5$ arcsec, and (ii) $i$-band magnitude of $m < 22.6$ for each of the two images.
    \item {Unresolved cases} are those where the angular separations between all the pairs are below $0.5$ arcsec and (ii) the combined blended light curve has peak brightness $m< 22.6$ in $i$-band.
\end{itemize}

Let us first focus on the macro-lensing-based calculation; we return to the estimations considering microlensing below.
Following the above definitions, we get $27.8$ resolved and $71.1$ unresolved LSNe Ia per cumulative year of LSST observation ---these numbers are marked by filled circles on the respective solid curves in Figure \ref{fig:Nstat}. Therefore, we expect to see $\sim 2.6$ times as many unresolved LSNe Ia in LSST as resolved lensed cases. Therefore, the detection of these unresolved cases is an important task.

In addition to introducing fluctuations in light curves, microlensing also either suppresses or boosts the brightness of images, while maintaining the same ensemble average as the macrolensing magnification. Intriguingly, microlensing tends to suppress brightness more frequently than boosting it \citep{Goldstein2018, Nikki_2023}, resulting in fewer systems being brighter than the critical magnitude ($m_{\rm crit}$) as compared to the estimations based on solely macrolensing magnification. This observation is evident as the dashed curves consistently lie below their macrolensed-only counterparts. However, since unresolved sources are permitted to be fainter than resolved sources (while both still need to be brighter than the limiting magnitude), the impact of microlensing effects on the number of unresolved lensed systems is less pronounced. In other words, summing up light curves of multiple images ---each affected differently by microlensing--- reduces the impact of microlensing on the peak brightness of the unresolved cases. Thus, the ratio of unresolved to resolved cases is further enhanced when one considers microlensing. For example, with $\theta_{\rm crit}=0.5$, we get approximately $21.5$ resolved and $68.2$ unresolved systems per cumulative year, boosting the ratio to $\sim 3.2$. However, while exploring the comparison of resolved versus unresolved systems in terms of multiple statistics we restrict our analysis to systems obtained using macro-magnification only.
Note that in the upcoming sections, where we aim to detect unresolved systems, we incorporate comprehensive microlensing effects into the light curves.

\begin{figure}
 \centering
\includegraphics[width=0.5\textwidth]{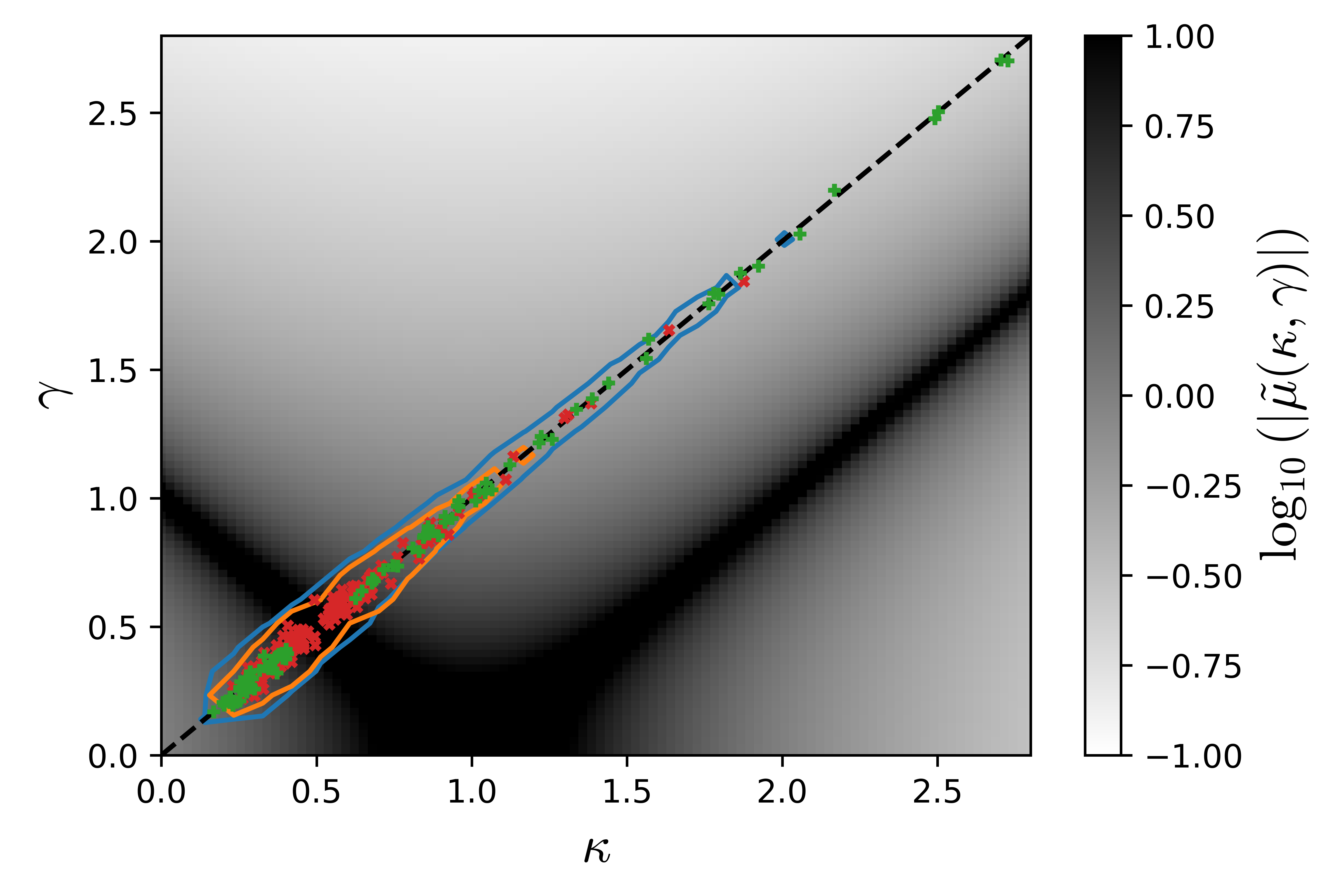}
\caption{The $95\%$ confidence level of the $(\kappa, \gamma)$ distribution compared for the images of resolved systems (orange contour) and unresolved systems (blue contour). The greyscale represents the macro-magnification in log-scale, clipped between $[0.1,10]$. Notably, the $(\kappa, \gamma)$ distribution for the unresolved systems extends towards higher values (compared to that of the resolved systems), accommodating lower magnifications. The crosses mark the $(\kappa, \gamma)$ values of the images used to simulate the microlensed light curves, with green and red crosses distinguishing images in doubles and quads, respectively. The diagonal dashed line denotes the $\kappa=\gamma$ line.}
\label{fig:kg_dist}
\end{figure}

Now, let us return to the forecast with macro-magnification-based calculation, which predicts $27.8$ resolved and $71.1$ unresolved LSNe Ia systems per effective year of LSST observation. We need to use the convergence and shear $(\kappa, \gamma)$ values at the LSN image positions of the unresolved systems when simulating their joint light curves with the microlensing effect, as the microlensing maps depend on these $(\kappa, \gamma)$ values.
 Figure \ref{fig:kg_dist} compares the $95 \%$ confidence levels (CLs) of the $(\kappa, \gamma)$ distributions for the unresolved (blue contour) and resolved samples (orange contour). Note that the requirement of lower $\theta$ for the unresolved cases does not affect the $(\kappa, \gamma)$ distribution as compared to that of the resolved cases. However, since unresolved sources are allowed to be fainter, they can accommodate lower magnifications and thus their $(\kappa, \gamma)$ distribution extends towards higher values. We use the distribution for the unresolved samples when simulating their microlensed light curves in the following section. We present the source and lens redshift distributions of the unresolved systems in Appendix \ref{app:unres_stat_extra}, comparing them against those of the resolved cases.

\begin{figure*}
 \centering
\includegraphics[width=0.537\textwidth]{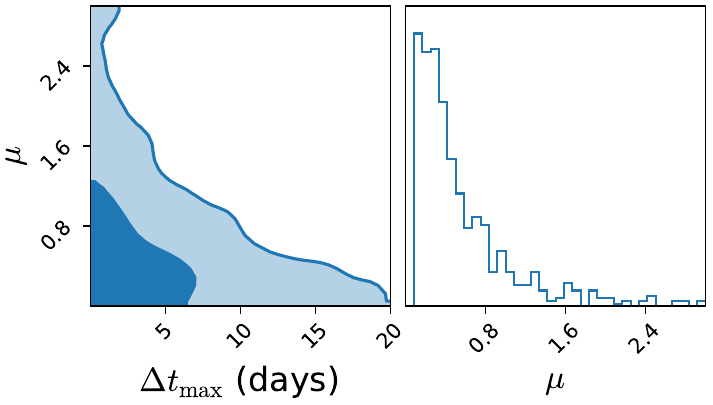}
\includegraphics[width=0.448\textwidth]{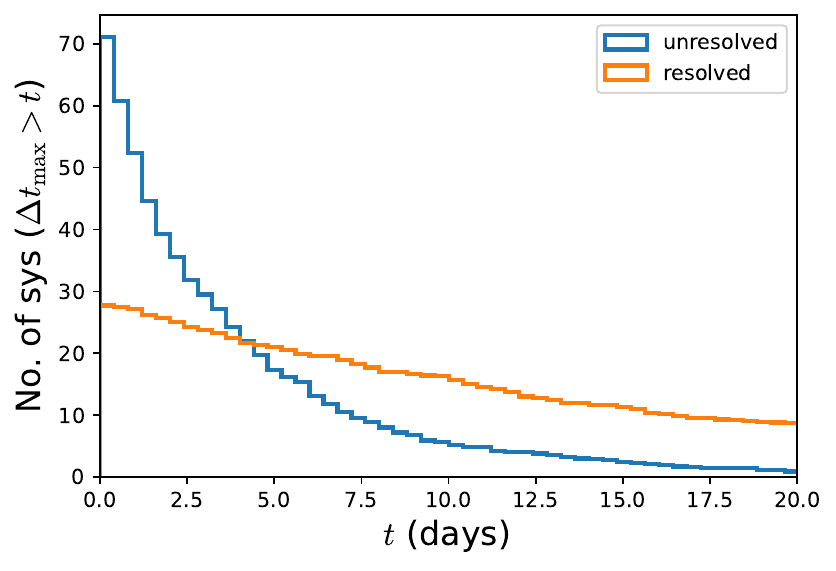}
\caption{Distribution of the largest time delay ($\dt_{\rm max}$) and the corresponding magnification ratio in all the unresolved systems is shown in the left panel; the lighter and darker shades of blue show the $68\%$ and $95\%$ credible regions, respectively. The middle panel presents the 1D marginalised distribution of the magnification ratio. The right panel shows the cumulative distribution of  $\dt_{\rm max}$, i.e. indicating the number of systems (in one cumulative year) that have $\dt_{\rm max}$ higher than a given value ---resolved (orange) vs unresolved (blue).
It is evident that unresolved systems typically exhibit low time delays, with a median value of $\dt_{\rm max}$ of $\sim 2.03$ days and approximately $10\%$ having values above $10$ days. }
\label{fig:td-stat}
\end{figure*}

Next, we compare the $\mu-\dt$ distribution for the resolved and unresolved cases in Figure \ref{fig:td-stat}. We remind the reader that we always sort the images chronologically and define the magnification ratios and time delays of the images with respect to the reference image, which arrives first in time (i.e. the first image). Consequently, a double system possesses only one set of $\mu$ and $\Delta t$, while a quad system entails three sets of $\mu$ and $\dt$. However, in the case of quads, the image with the largest time delay holds the greatest importance in the detection process. Therefore, we exclusively plot the $\mu$ and $\Delta t$ of the largest time-delay image for each quad system in Figure \ref{fig:td-stat}.

The left panel of Figure \ref{fig:td-stat} illustrates the $68\%$ and $95\%$ CLs of the $\mu-\dt$ distribution. The middle panel presents a 1D histogram of $\mu$ (specifically for the image with the largest time delay relative to the first image for quads). This histogram reveals a median $\mu$ of $0.40$, with a range of $(0.17, 0.61)$ at the $68\%$ percentile and $(0.11, 2.46)$ at the $95\%$ percentile.
Given that the first image, which acts as our reference, is usually the brightest for doubles and is brighter than the last image to appear for quads, the majority of the $\mu$ values fall below unity. The right panel displays the cumulative time-delay distribution, indicating how many systems have the largest time delay above a given $t$ and comparing it for resolved (orange) and unresolved (blue) samples. (We emphasise that the number of systems presented here corresponds to one cumulative year of observations, ensuring a fair and straightforward comparison between resolved and unresolved systems.)

As anticipated, unresolved systems generally exhibit smaller time delays compared to their resolved counterparts, rendering them less suitable for time-delay cosmography. For the unresolved systems, the median gives $\dt_{\rm max}=2.03$ days. Nevertheless, if a mechanism can be devised to detect the majority of unresolved systems, regardless of their small time delays, their sheer abundance could contribute significantly to achieving precision cosmology. Although detecting these low-time-delay systems solely through light-curve data is challenging, the aim of this article is to provide a method to accomplish this.

\section{Simulating mock blended joint light curves}
\label{sec:train_test_sim}

In this section, we detail the process of constructing the training, validation, and test sets comprised of the light curves of unresolved LSNe Ia ---consistent with the OM10 simulations--- and of unlensed SNe Ia, across multiple bands.

\subsection{Microlensed light curves}
In the previous section, we neglect microlensing when exploring the statistics of the forecasted unresolved LSNe Ia samples. Microlensing can affect the lensed light curves of individual images differently, and these effects may also differ across filters \footnote{However, multiple studies \citep[e.g.][]{Goldstein2018, holismokes3} have shown that there is an achromatic phase in the microlensing effects up to roughly three rest-frame weeks from the explosion.}. Therefore, we must take microlensing into account when training the models.

To calculate microlensed light curves of LSNe Ia, we need (i) specific intensity profiles, which we get from theoretical SN Ia models, and (ii) microlensing magnification maps. Details on the calculation of microlensed light curves are presented in \cite{Huber2017}.  As in \citet{Huber2017,2020A&A...644A.162S,holismokes3,holismokes7}, we use theoretical SN Ia models from \texttt{ARTIS} \citep{Kromer:2009ce} and magnification maps from \texttt{Gerlumph} \citep{Vernardos:2014lna,Vernardos:2014yva,Vernardos:2015wta,ChanEtal21}.  For the microlensing maps, we assume a Salpeter initial mass function (IMF),  as in \citet{holismokes7} and \citet{Huber2024}, with a mean mass of the microlenses of $0.35 M_\odot$ and a resolution of 20000 $\times$ 20000 pixels with a total size of 20 $R_\mathrm{ein}$ $\times$ 20 $R_\mathrm{ein}$ on the source plane.

Another important factor for generating microlensing maps is the smooth matter fraction ($s \in [0,1]$) that determines what fraction of the total matter is smooth while the rest is in the form of stars at an image location; lower $s$ would lead to more severe microlensing modifications to the macrolensed light curve of an image. The saddle images, which often lie closer to the lens centres especially for doubles, tend to have lower smooth-matter fractions. Conversely, the minima images often lie further away from the lens centres and thus typically have higher values of $s$. Therefore, to maintain a balance between realism and simplicity, we assign $s=0.4$ for all the saddle images and randomly choose either $0.6$ or $0.8$ for the minima images.

Finally, we simulate the microlensed light curves for a total of $250$ images, stemming from randomly selected $49$ doubles and $38$ quads. The $\kappa, \gamma$ values of these $250$ images are shown by cross symbols in Figure \ref{fig:kg_dist};  with green and red crosses indicating images of doubles and quads, respectively. As evident from the figure, the $(\kappa, \gamma)$ values of the downsampled images adequately represent the overall distribution of these parameters for all unresolved systems. For a given image, we generate a microlensing map using its $(\kappa, \gamma, s)$ values. Subsequently, we simulate $100$ microlensing realisations corresponding to $100$ random locations on the microlensing map. The selection of the SN Ia model can also play an important role. To maintain a model-agnostic approach, we should incorporate a mixture of different SN Ia models during the training process. Among the four theoretical models investigated in \citet{holismokes7}, the merger model \citep{Parkmor2011,Parkmor2012} exhibits the most diverse behaviour, the other three being W7 (carbon deflagration, \citet{Nomoto1984}), N100 (delayed detonation, \citet{Ropke2012, Seitenzahl2013}), and sub-Ch (sub-Chandrasekhar mass detonation, \citet{Sim2010}). For simplicity, while still retaining diversity in the SN Ia models, we opt to randomly choose between the W7 and merger models when simulating the light curves for each system.

When we construct the blended light curve of a lensed unresolved SN, we randomly select a system from the corresponding sample (double or quad) with microlensing simulations available. We then randomly choose one microlensing realisation for each of the images and sum up the microlensed light curves separately in each filter.

\subsection{Unlensed light curves}
While simulating the light curves for an unlensed SN Ia, we randomly select one of the lensed systems for which the simulated light curves are available. Subsequently, we obtain the intrinsic light curve by dividing the macrolensed light curve of any of the images by the corresponding macro-magnification.

\subsection{LSST cadence distribution and noise implementation}
\label{sec:lsst_cadence_noise}

\begin{table*}
\caption{
Statistics from observing strategy {\tt baseline v3.2} in the four LSST filters: $r$, $i$, $z$, and $y$. }
\resizebox{\textwidth}{!}{\begin{tabular}{|c|c|c|c|c||c|c|c|c|}
  \cline{2-9}
\multicolumn{1}{c|}{} &  \multicolumn{4}{c||}{\makecell{Systems individually well-observed in 4 bands \\ (4 sets)}} & \multicolumn{4}{c|}{\makecell{Systems well-observed  in all 4 bands \\ (intersection of the 4 sets)}} \\
\hline
 Bands & $r$ & $i$ & $z$ & $y$  & $r$ & $i$ & $z$ & $y$ \\
  \hline
  \hline
 \makecell{Good \\ systems} &  \makecell{$13.0\%$ } & \makecell{$14.0 \%$} & \makecell{$12.2 \%$}& \makecell{$9.8 \%$} &  \multicolumn{4}{c|}{6.2 \%}\\
 \hline
    \makecell{Avg no. \\ of  epochs}  &\makecell{ $ 21.4$} & \makecell{ $21.6 $ } & \makecell{  $20.8 $ } & \makecell{ $19.9$ } & $24.9 $ & $26.1 $ & $ 24.2$ & $ 22.3$\\
  \hline
  \makecell{Cadence \\ mean \& median \\ (days)}  &  \makecell{ $ 7.3 ~\&~ 4.0$ } & \makecell{ $7.6~\&~5.0 $} & \makecell{  $8.4 ~\&~ 5.7$ } & \makecell{ $11.6 ~\&~ 5.8$ } & $ 6.7 ~\&~ 4.0 $ & $6.7 ~\&~ 4.8 $ & $ 7.6 ~\&~ 5.0$ & $ 10.8 ~\&~ 5.0$\\
  \hline
  \makecell{Largest gap \\ mean \& median \\ (days)} & \makecell{ $ 28.3 ~\&~ 24.9$ } & \makecell{ $27.9 ~\&~23.9 $} & \makecell{  $29.1 ~\&~ 25.8$ } & \makecell{ $33.3 ~\&~ 28.9$ } & $ 24.5 ~\&~ 23.0$ & $23.9 ~\&~ 21.9 $ & $25.9 ~\&~ 23.9 $ & $ 31.6 ~\&~ 28.8$\\
  \hline
 \makecell{ Mean $m_5$} & $ 23.9 $ & $23.3 $ & $ 22.7$ & $21.9 $ & \multicolumn{4}{c|}{same as in the individual 4 sets corresponding to 4 bands}\\
  \hline
\end{tabular}}
\tablefoot{We have a set of `good systems' that meet the criteria for `good observation', as described in Section \ref{sec:lsst_cadence_noise}, for each of these four filters. The table is divided into two parts, each with four columns. The left part presents various statistics, derived from the observing strategy, considering these four sets of systems well observed in the four bands. The right half presents the same statistics but for the systems well observed in all four bands, i.e., for the intersection of the four sets. For example, the left part of the second row shows the fraction of unresolved systems that would be well observed in different bands, whereas the right part indicates that $6.2\%$ of the systems will be well observed in all four bands. Note that, in comparison, $\sim 40\%$ of the total resolved systems will be detected due to season gaps in the baseline observing strategy \protect\citep{Huber2017}. The next three rows present the average number of observation epochs, the mean and median cadence, and the largest gap in it, respectively, in the four bands considering the four sets of systems individually in the left part and their intersection in the right part. The last row provides the mean 5$\sigma$ depth ($m_5$) in the four bands; these numbers should remain the same for the intersection part as well. }
\label{tab:os_3.2}
\end{table*}

We implement a realistic LSST cadence using publicly available simulations from the {\it Rubin} Operations Simulator (\texttt{OpSim}), which emulates the field selection and image acquisition process of LSST over its 10 year duration \citep{Delgado2014, Delgado2016, Naghib2019}. The main survey mode of LSST is the Wide-Fast-Deep (WFD) programme, which will cover around 18,000 square degrees, corresponding to $\sim 90\%$ of the observing time. Following the latest recommendations from the Survey Cadence Optimization Committee\footnote{\href{https://pstn-055.lsst.io/}{https://pstn-055.lsst.io/}}, the WFD survey will employ a rolling cadence, in which alternating areas of the sky will receive more frequent visits than others. As a result, objects discovered in the high-cadence region will have more observations and, hence, a better light-curve coverage. In our simulations, we consider LSNe Ia in the WFD region as well as in the remaining sky regions: the Galactic plane, polar regions, and the Deep Drilling Fields, encompassing a total footprint of 20,000 square degrees. We implement the \texttt{baseline v3.2} observing strategy to simulate our mock light curves. Among the six LSST filters, we consider only `$r$', `$i$', `$z$', and `$y$' bands, as the remaining two, namely `$u$' and `$g$' bands, have much `lower' cadences.

To sample the (joint) light curve according to the observing strategy, we randomly choose $7\,110\,000$ sky locations (which corresponds to an oversampling of $100\,000$ times the number of unresolved SNe per cumulative year of LSST observations) within the LSST footprint. We use \texttt{OpSimSummary}~\citep{Biswas2020} to retrieve all observation times, filters, and 5$\sigma$ depths corresponding to a specific sky location within the field of view of the telescope.
For each sky location, we then randomly choose a time within the 10 year run of LSST that represents the epoch of SN explosion. Many of these SNe will remain unobserved due to season gaps and some will be observed only for a few epochs, leading to partially monitored light curves. To filter out the suitable systems that can have light curves observed for sufficient epochs, we apply the following criteria,:
\begin{enumerate}
    \item Number of observations of $> 10$ within $120$ days from the explosion.
    \item First observation within $20$ days of the explosion.
    \item At least one observation after $40$ days post-explosion.
\end{enumerate}

When a SN satisfies the above criteria in a filter, we consider that the system has `good' observation in that particular filter. Therefore, we have four sets of well-observed systems in the four LSST filters that we consider: $r$, $i$, $z,$ and $y$. Various statistics for these sets, derived from the observation strategy, are presented in Table \ref{tab:os_3.2} in two parts (each with four columns corresponding to four bands). The left part considers the four sets of good systems individually, whereas the right part presents the same statistics but for the subset of the systems that are well-observed in all four bands, that is, for the intersection of the four sets. For example, the second row presents the percentage of the unresolved systems that will be well observed in different bands in the left part, whereas the right part informs us that $6.2\%$ will be well observed in all four bands. The union of good systems covers $18.8\%$ of the total sample. The third row shows the mean number of observed epochs (up to $120$ days from the explosion) in different bands. The next two rows  present the distributions of the cadence and the largest gap in observation epochs, respectively, in terms of the mean and median of the respective quantities. The last row indicates the mean 5$\sigma$ depth in different bands; these numbers should remain the same for the intersection set as well.

Overall, we find that $r$ and $i$ bands provide better cadence distributions and smaller maximum gaps in observational epochs, as compared to the other two remaining bands. Therefore, the number of `good systems' is higher in these two ($r$ and $i$) bands. Fortunately, SNe Ia exhibit higher S/N in these two bands due to their higher values of 5$\sigma$ depth ($m_5$), making them ideal for this study.

\subsection{Uncertainties}
The {\tt Baseline v3.2} observing strategy provides the 5$\sigma$ depth ($m_5$) in each filter at every epoch. The bottom row of Table \ref{tab:os_3.2} presents the mean $m_5$ in each filter.
Following \citet{lsst1}, we simulate the uncertainties on the magnitudes as
\beq
\sigma_{\rm tot}^2=\sigma_{\rm sys}^2+\sigma_{\rm rand}^2
,\eeq
where we set the systematic noise to $\sigma_{\rm sys}=0.005$ (the upper limit) for all filters, the random noise is given by
\beq
\sigma_{\rm rand}^2=(0.04-\gamma)x + \gamma x^2~,~{\rm where}~ x=10^{m-m_5}
,\eeq
and the filter-dependent quantity $\gamma= 0.039$ for $r$ and $i$ bands and $0.040$ for $z$ and $y$ bands; see \citet{lsst1} for details.

\begin{figure*}
 \centering
 \subfigure[Light curves in 4 bands]{
\includegraphics[width=0.477\textwidth]{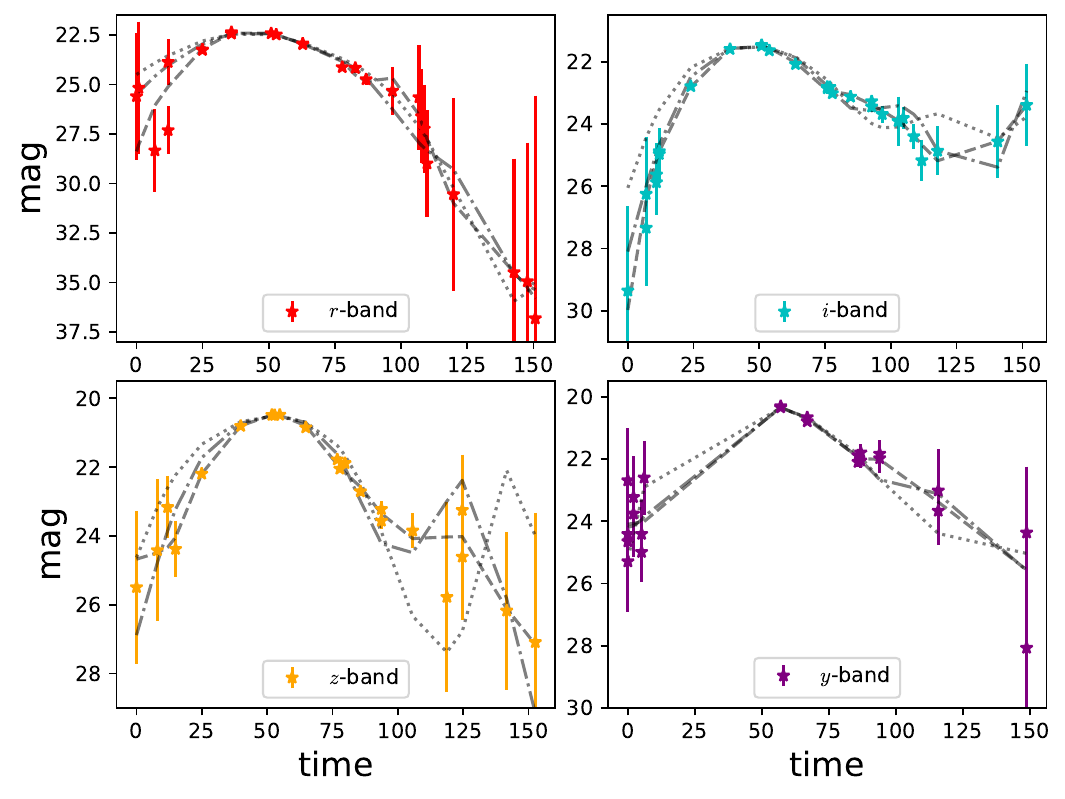}\label{fig:lc_inp_a}}
\subfigure[1D input to the CNN in 3 channels (shown in vertical panels)]{
\includegraphics[width=0.493\textwidth]{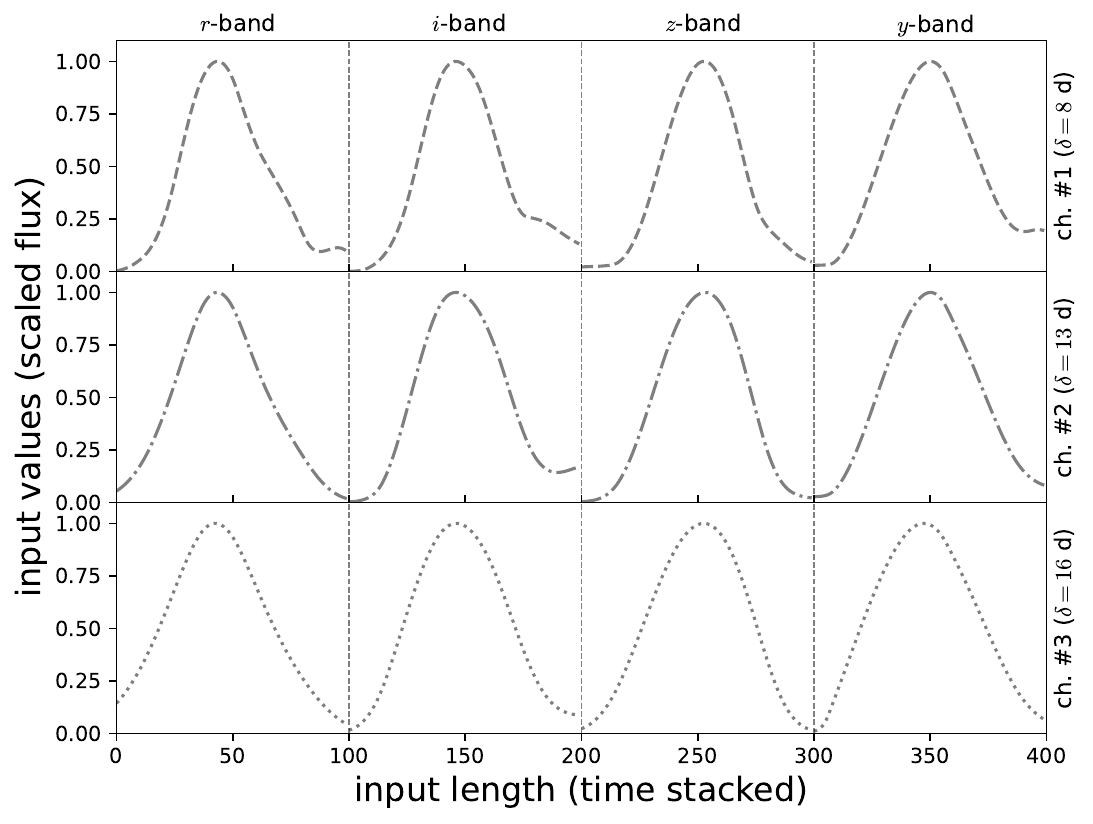}\label{fig:lc_inp_b}}
\caption{{\bf(a)} Blended light curves of a typical unresolved system are shown in the left panel across four LSST bands. We adopt LSST-like cadence and noise to generate mock light curves (as detailed in Sec. \ref{sec:lsst_cadence_noise}). Data processing begins by smoothing these light curves using different smoothing scales. The grey curves, delineated with dashed, dash-dotted, and dotted line styles, represent the smoothed light curves corresponding to smoothing scales of 8, 13, and 16 days, respectively. Subsequently, we convert the smoothed magnitudes into fluxes and sample the light curves on a predefined set of epochs, kept fixed for each band. For simplicity, we opt for 100 epochs with 1 day spacing, starting from the initial observation epoch (set as $t=0$), for all the bands. We then normalise every light curve separately by scaling the fluxes in the range $[0,1]$.
{\bf (b)} Finally, for each smoothing scale, we stack the normalised flux-versus-time light curves corresponding to the four bands side by side, thereby recasting the data into an array of 400 in  length. Subsequently, we arrange three such arrays ---corresponding to the three choices of smoothing scale--- into three channels. Consequently, the final data size for each system becomes $(400,3)$, as depicted in the right panel. This sample of data, of fixed size, is then fed into the CNNs.}
\label{fig:lc_inp}
\end{figure*}

Note that when the true (or theoretical) magnitude is much larger than $m_5$, that is, the SN light curve is too faint compared to the 5$\sigma$ depth, the $\sigma_{\rm rand}$ can be quite high. In these cases, a simple normal distribution can often incorrectly produce very bright observations (significantly brighter than the $m_5$ itself). To avoid this, we generate the observed magnitude by randomly choosing a value $m \sim {\rm Uniform} (m_5,m_{\rm true})$ (see \cite{holismokes7}); however this is of little importance as these are typically the faintest parts of the SN light curves. The left panel of Fig. \ref{fig:lc_inp} presents the unresolved light curves of a typical quad system in four bands as an example.

\section{Methodology: One-dimensional CNN}
\label{sec:method}
Inspired by \citet{misha2022}, we also employ 1D CNNs to detect the lensed cases from the light curve data.

\subsection{Data processing}
\label{sec:data_preprocess}
In reality, we do not know the explosion epoch of any LSN. To avoid providing any information regarding the explosion time to the network, we set the time to zero at the epoch of the first observation of each light curve.
The SNe Ia light curve sample from LSST will contain a different number of observations per system, as well as different observation epochs. Nevertheless, to be able to employ a simple 1D CNN on the mock observed light curves we need the input to the networks to have the same dimensions. To achieve this, in addition to considering observation noise, we adopt the following data processing.
\begin{itemize}
    \item We first smooth the observed magnitude versus time light curves using the iterative smoothing process of \cite{Shafieloo:2005nd}, which takes the uncertainties into account. Then we convert the smoothed magnitudes into the fluxes (in arbitrary units) to get flux versus time light curves. Next we interpolate (cubic spline) the smoothed light curves and extrapolate them at the tails if necessary to have the same data length; for example, $N_j$ for the $j$-th filter. Next we normalise the light curves by scaling all the fluxes from $0$ to $1$ (for all systems, separately in all filters). At first, we want to test the method on long light curves by empirically choosing $N_j=100$ for all the four filters, which means we use light-curve data until $100$ days from the first observation for all the systems across all bands. Later in the article, we reduce $N_j$ while trying to detect the lensed system early enough so that the SNe Ia could still  be visible in the follow up.

    \item We then concatenate (i.e. stack side by side) the normalised smoothed light curves for different bands (if available) to have an 1D array of $N_{\rm tot}=\sum_{j \in \rm filters}^{4} N_j$ data points comprising light curves for all the four bands. Note that we test the classification results using both stacked multi-band light curves and single-band observations.

    \item Furthermore, we use three different smoothing scales -- $8, 13,$ and $16$ days -- and pass the 1D processed data (each having $N_{\rm tot}$ length) to the CNNs in three different channels. This choice of the smoothing scale is consistent with the cadence distribution (considering the mean value and the gaps) and is kept the same across all filters. \footnote{The selection of the smoothing scales is important because larger scales can wipe out small-scale features, if present, while smaller scales may overfit and struggle with long gaps in the observation epochs. We tested multiple combinations of smoothing scales and found this choice to perform the best for the LSST cadence.}  The smoothed magnitude versus time light curves, corresponding to the three smoothing scales, are shown in each of the four panels of Figure \ref{fig:lc_inp_a}.
    We expect each channel to provide features at three different timescales, if any, in the smoothed light curves.

    \item Therefore, at the end, our input data eventually take the shape $(N_{\rm tot},3)$. Figure \ref{fig:lc_inp_b} illustrates how we arranged the input data for the example quad system. The three rows show the three 1D input arrays, each having a length of $N_{\rm tot}=400$, corresponding to three different smoothing scales, $\delta =8 , 13$, and  $16$ days.
\end{itemize}
In addition to accounting for different numbers of observational epochs for different systems, this choice of data preprocessing handles observation noise, which would influence the final smoothed light curves at different smoothing scales.

\subsection{Training and test sets}
\label{sec:train_data}

\begin{table}
  \caption{Description of different sets used for training and testing. }
  \centering
  \begin{tabular}{cccc}
    \hline
    Set name & $\dtt$ & \makecell{Use in \\classification} & \makecell{Use in \\ $\dt$ estimation} \\
    \hline
    set A & $U[5.0,24.0]$ & \multirow{3}{*}{ train + test} & \multirow{3}{*}{test only} \\
    set B & $U[4.0,24.0]$& &  \\
    set C &  $U[3.0,24.0]$& & \\
    \hline
    set D &$U[0.5,30.0]$& -- & train only\\
    \hline
    set LSST & LSST like & test only & test only\\
    \hline
  \end{tabular}
  \tablefoot{Note that, except for set LSST, in the rest of the controlled sets we use $\mu \sim U[1/3,3.0]$ to have a balanced training.
  Set D is only used to train the model for predicting time delays. Set LSST represents the whole sample of unresolved systems in LSST without any cuts being imposed. We use this set to test our models trained on different sets depending on our purpose of classification or time-delay estimation. Note that we randomly choose between W7 and merger models while generating the light curves for each system (lensed or unlensed) in all the sets. }
  \label{tab:set_abc}
\end{table}

Figure \ref{fig:td-stat} shows the distribution of maximum time delay and the corresponding magnification ratio for the unresolved systems in LSST. However, instead of using these distributions, we use simple uniform distributions for $\mu$ and $\dt$ to construct balanced training sets.
As explained in Section \ref{sec:1}, low values of time delays $(\dt \to 0)$ or extreme values of magnification ratios ($\mu \gg 1$ or $\mu \ll 1$) make our goal of detecting the lensed systems using blended joint light curves challenging.

At first, we set a reasonable limit on the magnification ratio, $\mu \sim U[1/3,3.0]$. However, to test the working limits of this approach in terms of the minimum feasible time delay, we constructed three different controlled sets, each with a different minimum $\dt$: (i) $\dt \in U[5.0,24.0]$ for set A,  (ii) $\dt \in U[4.0,24.0]$ for set B, and (iii) $\dt \in U[3.0,24.0]$ for set C. These sets are summarised in Table \ref{tab:set_abc}. We also created a separate test set (along with the validation set) for each training set with the same distributions of $\mu$ and $\dt$ for initial tests. We keep a ratio of $0.5:0.25:0.25$ between the population of unlensed, doubles, and quads, respectively, in all training and test sets used for classification.

 Note that even when the macro-magnification ratio, $\mu$, is non-extreme (e.g. around unity), microlensing can stochastically make it extreme in some cases. Inclusion of these joint light curves (comprising extreme values of $\mu$) in the training set can confuse the training process, as the light curves can be very similar to an unlensed case. This can eventually result in lower accuracy in classification (as well as time-delay estimations). Therefore, we always scale the microlensed light curves of the images of a system such that their $\mu$ match the input values (drawn from a uniform distribution), and the $i$-band peak brightness matches the corresponding value in the OM10 simulation. By scaling with a constant factor, we preserve the microlensing effects (i.e. those that lead to wiggles) on each light curve. However, although microlensing should roughly preserve the ensemble average of $\mu$, we can expect some cases to have extreme $\mu$ in reality due to microlensing. We anticipate that these systems will simply be classified as unlensed cases by the model, which is trained on a set that excludes extreme $\mu$.

While predicting the time delays for a test set, if we use the model trained on the corresponding training set with the same $\dt$ interval, say $\dt \in (\dt_{\rm min},\dt_{\rm max})$, we will inevitably face bias in the results at the $\dt$ boundaries. In this case, the model tends to over(under)-predict for smaller(larger) time delays. This is simply because the model will not learn to predict values lower than $\dt_{\rm min}$ and higher than $\dt_{\rm max}$. To overcome this apparent bias issue, we train the model on systems with $\dt \sim U[0.5,30.0]$ days (set D, which covers a larger range of $\dt$ than sets A, B, and C) to predict the time delays for the test sets -- A, B, and C. In this training set for time-delay estimation, we keep the populations of doubles and quads the same. Note that the test sets also share the same ratio.

Finally, to assess how our models ---which are trained on the above sets--- perform in realistic scenarios, we test them on a set with LSST-like unresolved systems (set LSST) without any restrictions on $\mu$ and $\dtt$.

\subsection{Network architectures}
As mentioned above, this work focuses on two distinct goals: (i) classification, that is, discerning the lensed systems from the unlensed ones (if possible classifying them into unlensed, doubles, and quads) and (ii) estimation of the time delays. Therefore, we employ two types of 1D CNN designed for the above two purposes: classification and time-delay regression. However, for simplicity, we keep the architecture of the two types of network quite similar until the last layer -- starting with two 1D convolution layers (each is followed by a max-pooling layer) followed by two dense layers.\footnote{When we tried to optimise the architecture parameters (such as the  number of convolution layers, filter sizes and numbers, and the number of dense layers) using {\tt keras.tuner,} keeping the overall structure the same, we could only find minute improvements. Therefore, for simplicity, in this article we keep the architecture parameters the same in the two types of network until the last layer.} The last layer is adapted according to the purpose of classification or regression. We emphasise that we only use simple network architectures in this work; however, a completely different (e.g. long-short term memory or LSTM) network can result in better results. Exploring all possible network architectures is beyond the scope of this work, the aim of which is simply to demonstrate the overall feasibility of classifying and analysing unresolved SN light curves with CNNs, but this could be an interesting avenue to explore in a follow-up project.

\section{Results}

\subsection{Classification using long light curves}
\label{sec:full_lc}
Initially, we use the long light curves, until $100$ days from the first observation in each band.
\subsubsection{$3 \times 3$ classification: unlensed versus doubles versus quads}

\begin{table}
  \caption{Confusion matrices for the classification of unlensed, double, and quad systems.}

  \begin{minipage}{\linewidth}
    \centering
\begin{tabular}{ccc|c|c|}
\multicolumn{2}{c}{} & \multicolumn{3}{c}{\textbf{Predictions}} \\ \cline{3-5}
\multicolumn{2}{c|}{} & \textbf{unlensed} & \textbf{double} & \textbf{quad} \\  \cline{2-5}
\multirow{5}{*}{\rotatebox[origin=c]{90}{\textbf{Truth}}} &\multicolumn{1}{|c|}{\textbf{unlensed}} & \makecell{$ \mathbf{all: 98.4 \%}$ \\  $r$: $  93.8\% $  \\  $i$: $  96.8\% $ } & \makecell{$\mathbf{all: 1.3 \%}$  \\ $r$: $  4.7\% $ \\ $i$: $  2.6\% $ } & \makecell{$\mathbf{all: 0.3 \%}$  \\ $r$: $  1.5\% $ \\ $i$: $  0.6\% $ } \\ \cline{2-5}

& \multicolumn{1}{|c|}{\textbf{double}} & \makecell{$\mathbf{all: 14.0\%}$  \\ $r$: $  25.8\% $  \\ $i$: $  20.2\% $ } & \makecell{$\mathbf{all: 70.2\%}$ \\ $r$: $  54.5\% $ \\ $i$: $  55.1\% $ } & \makecell{$\mathbf{all: 15.8\%}$ \\ $r$: $  19.7\% $  \\ $i$: $  24.7\% $ } \\ \cline{2-5}

& \multicolumn{1}{|c|}{\textbf{quad}} & \makecell{$\mathbf{all: 2.4\%}$  \\ $r$: $  9.2\% $ \\ $i$: $  8.2\% $ } & \makecell{$\mathbf{all: 9.9\%}$  \\ $r$: $  18.5\% $ \\ $i$: $  19.5\% $ } & \makecell{$\mathbf{all: 87.7\% }$  \\ $r$: $  72.3\% $ \\ $i$: $  72.3\% $ } \\ \cline{2-5}

\end{tabular}\\

\vspace*{2mm}
\centering
{set A: $\dtt \sim  U[5.0,24.0]~{\rm and}~ \mut \sim U[1/3,3.0]$}
\vspace*{3mm}
  \end{minipage}%

  \begin{minipage}{\linewidth}
\centering
\begin{tabular}{ccc|c|c|}
\multicolumn{2}{c}{} & \multicolumn{3}{c}{\textbf{Predictions}} \\ \cline{3-5}
\multicolumn{2}{c|}{} & \textbf{unlensed} & \textbf{double} & \textbf{quad} \\  \cline{2-5}
\multirow{5}{*}{\rotatebox[origin=c]{90}{\textbf{Truth}}} &\multicolumn{1}{|c|}{\textbf{unlensed}} & \makecell{$\mathbf{all: 97.0\%}$ \\  $r$: $  93.9\% $  \\  $i$: $  96.6\% $ } & \makecell{$\mathbf{all: 2.6\%}$ \\ $r$: $  4.2\% $  \\ $i$: $  2.7\% $  } & \makecell{$\mathbf{all:  0.4\%}$ \\ $r$: $  1.9\% $  \\ $i$: $  0.7\% $ } \\ \cline{2-5}

& \multicolumn{1}{|c|}{\textbf{double}} & \makecell{$\mathbf{all: 14.0\%}$ \\ $r$: $  28.3\% $  \\ $i$: $  21.9\% $ } & \makecell{$\mathbf{all: 69.6\%}$  \\ $r$: $  52.8\% $  \\ $i$: $  52.0\% $} & \makecell{$\mathbf{all: 16.4\%}$  \\ $r$: $  18.9\% $ \\ $i$: $  26.1\% $} \\ \cline{2-5}

& \multicolumn{1}{|c|}{\textbf{quad}} & \makecell{$\mathbf{all: 3.4\%}$ \\ $r$: $  10.3\% $ \\ $i$: $  9.6\% $ } & \makecell{$\mathbf{all: 11.1\%}$ \\ $r$: $  19.5\% $ \\ $i$: $  19.5\% $ } & \makecell{$\mathbf{all: 85.5\% }$ \\ $r$: $  70.2\% $ \\ $i$: $  70.9\% $ } \\ \cline{2-5}

\end{tabular} \\
\vspace*{2mm}
\centering
{set B: $\dtt \sim  U[4.0,24.0]~{\rm and}~ \mut \sim U[1/3,3.0]$}
\vspace*{3mm}

  \end{minipage}%


  \begin{minipage}{\linewidth}
  \centering
\begin{tabular}{ccc|c|c|}
\multicolumn{2}{c}{} & \multicolumn{3}{c}{\textbf{Predictions}} \\ \cline{3-5}
\multicolumn{2}{c|}{} & \textbf{unlensed} & \textbf{double} & \textbf{quad} \\  \cline{2-5}
\multirow{5}{*}{\rotatebox[origin=c]{90}{\textbf{Truth}}} &\multicolumn{1}{|c|}{\textbf{unlensed}} & \makecell{$\mathbf{all: 95.2\%}$ \\  $r$: $  93.4\% $ \\  $i$: $  96.0\% $ } & \makecell{$\mathbf{all:  3.9\%}$ \\ $r$ : $  4.7\% $  \\ $i$: $  2.8\% $ } & \makecell{$\mathbf{all:  0.9\%}$  \\ $r$: $  1.9\% $ \\ $i$: $  1.2\% $ } \\ \cline{2-5}

& \multicolumn{1}{|c|}{\textbf{double}} & \makecell{$\mathbf{all:  16.8\%}$ \\ $r$: $  30.0\% $ \\ $i$: $  24.2\% $ } & \makecell{$\mathbf{all:  65.0\%}$ \\ $r$: $  49.1\% $ \\ $i$: $  47.9\% $ } & \makecell{$\mathbf{all:  18.2\%}$  \\ $r$: $  20.9\% $ \\ $i$: $  27.9\% $ } \\ \cline{2-5}

& \multicolumn{1}{|c|}{\textbf{quad}} & \makecell{$\mathbf{all:  3.6\%}$  \\ $r$: $  12.0\% $ \\ $i$: $  9.8\% $ } & \makecell{$\mathbf{all:  11.8\%}$ \\ $r$: $  17.4\% $  \\ $i$: $  17.6\% $ } & \makecell{$\mathbf{all:  84.6\% }$ \\ $r$: $  70.6\% $ \\ $i$: $  72.6\% $  } \\ \cline{2-5}

\end{tabular}\\
\vspace*{2mm}
\centering
{set C: $\dtt \sim  U[3.0,24.0]~{\rm and}~ \mut \sim U[1/3,3.0]$}
  \tablefoot{The results obtained from multi-band ($rizy$) light curves are labelled `all,' while those from individual $r$ and $i$ bands are labelled accordingly.}
  \label{tab:cm_33}
  \end{minipage}

\end{table}

\begin{table}
\caption{Set A: $\dtt \sim  U[5.0,24.0]~{\rm and}~ \mut \sim U[1/3,3.0]$}
\centering
\begin{tabular}{ccc|c|}
\multicolumn{2}{c}{} & \multicolumn{2}{c}{\textbf{Predictions}} \\ \cline{3-4}
\multicolumn{2}{c|}{} & \textbf{unlensed} & \textbf{lensed} \\  \cline{2-4}
\multirow{4}{*}{\rotatebox[origin=c]{90}{\textbf{Truth}}} &\multicolumn{1}{|c|}{\textbf{unlensed}} & \makecell{$ \mathbf{all:  97.1\%}$ \\  $r$: $  92.9\% $ \\  $i$: $  95.8\% $ \\  $z$: $  93.5\% $\\  $y$: $  86.1\% $ } & \makecell{$ \mathbf{all:  2.9\%}$ \\  $r$: $  7.1\% $ \\  $i$: $  4.2\% $ \\  $z$: $  6.5\% $\\  $y$: $ 15.9 \% $ }  \\ \cline{2-4}

& \multicolumn{1}{|c|}{\textbf{lensed}} & \makecell{$ \mathbf{all:  7.7\%}$ \\  $r$: $  16.6\% $ \\  $i$: $  12.5\% $ \\  $z$: $  17.7\% $\\  $y$: $  25.1\% $ } & \makecell{$ \mathbf{all:  92.3\%}$ \\  $r$: $  83.4\% $ \\  $i$: $  87.5\% $ \\  $z$: $  82.3\% $\\  $y$: $  74.9\% $ }  \\ \cline{2-4}
\end{tabular}
\label{tab:cm22_dtmin_5}
\end{table}

First we tried to classify the three types of systems -- unlensed, doubles, and quads. The results are summarised in Table \ref{tab:cm_33} for the three sets (A, B, and C in Table \ref{tab:set_abc}) with minimum $\dt$ being $5.0$, $4.0,$ and $3.0$ days from the top. The first row of a confusion matrix provides the percentage of the true unlensed systems that are classified as unlensed, double, and quads in the three respective columns; similarly, the next two rows provide the classification percentages of true doubles and quads. In each cell, we provide the result for all band data, considering only $r$-band, and only $i$-band data. As expected, the classification results are the best when the multi-band data are used; however, $r$ or $i$ band only data also result in reasonably good classification matrices.

  As the lensing probability is low \citep[as we expect only one lensed system out of $\sim10^3 - 10^4$ randomly chosen SNe Ia;][]{om10}, we need to control the false-positive rate (FPR) very well. For example, if the lensing probability is $0.01\%$, a FPR of $10\%$ gives roughly $1000$ false alarms per one true detection, assuming the perfect true-positive rate (TPR) of $100\%$.
We find that for set A (with $\dt_{\rm min}=5.0$ days), the networks can classify with a FPR of as low as $\sim 1.6 \%$.
However, the FPR increases to $\sim 4.8\%$ for set C, where $\dt_{\rm min}$ is $3$ days. The results mentioned above appears slight inferior as compared to those presented in \citet{misha2022}, due to differences in cadence, uncertainties, and the distribution of $(\mu,\dt)$ between LSST observation conditions considered here and the setup employed in \citet{misha2022}.

 Table \ref{tab:cm_33} (the four bottom-right cells) clearly shows  that the network confuses between doubles and quads very often. This confusion arises because we do not keep any minimum separation between the time delays of the later three images (with respect to the first image) for the quads. As a consequence, the joint light curve of a quad with three images with similar time delays, say $\dt '$ from the first image, will be similar to a double with the time delay $\dt '$ but with an aggregated magnification. Therefore, the network is expected to confuse between certain quads and doubles. Also, owing to this confusion,
the FPRs in Table \ref{tab:cm_33} exhibit a variation of $\sim$$1$-$2\%$ (additional variation on the FPR, not percentage of FPR) due to stochasticity in the network training, that is, with different network initialisations. Therefore, it is more logical to classify the systems as unlensed versus lensed.

\subsubsection{$2 \times 2$ classification: lensed versus unlensed}
Here we consider that all doubles and quads have the same truth label -- `lensed'. The classification results are presented in Tables \ref{tab:cm22_dtmin_5}, \ref{tab:cm22_dtmin_4},
 and \ref{tab:cm22_dtmin_3} for the three sets -- A, B, and C. The top number (in boldface) in each cell corresponds to the scenario with complete four-band light-curve data. The subsequent four numbers are obtained using single-band light-curve data, as labelled. The TPR remains at $\gtrsim 90\%$, with an FPR of around $3\%$, for all three sets when we consider the four-band data. However, excluding the lower time-delay systems (going from set C to A) leads to enhanced overall accuracy, as expected.

Note that, as opposed to the $y$-band, the three bands $r$, $i$, and $z$ individually produce well-controlled FPR of approximately $4-7\%$, while maintaining acceptable TPR of approximately $80-92\%$. However, it is evident that $i$-band light curves provide the most reliable classification because of the improved cadence and S/N. Therefore, even with single-band light-curve data from any of these three bands, we can achieve reasonably accurate classification.

\begin{table}
\caption{Set B: $\dtt \sim  U[4.0,24.0]~{\rm and}~ \mut \sim U[1/3,3.0]$}
\centering
\begin{tabular}{ccc|c|}
\multicolumn{2}{c}{} & \multicolumn{2}{c}{\textbf{Predictions}} \\ \cline{3-4}
\multicolumn{2}{c|}{} & \textbf{unlensed} & \textbf{lensed} \\  \cline{2-4}
\multirow{4}{*}{\rotatebox[origin=c]{90}{\textbf{Truth}}} &\multicolumn{1}{|c|}{\textbf{unlensed}} & \makecell{$ \mathbf{all:  96.8\%}$ \\  $r$: $  91.6\% $ \\  $i$: $  95.2\% $ \\  $z$: $  92.6\% $\\  $y$: $  83.8\% $ } & \makecell{$ \mathbf{all:  3.2\%}$ \\  $r$: $  8.4\% $ \\  $i$: $  4.8\% $ \\  $z$: $  7.4\% $\\  $y$: $  16.2\% $ }  \\ \cline{2-4}

& \multicolumn{1}{|c|}{\textbf{lensed}} & \makecell{$ \mathbf{all:  8.9\%}$ \\  $r$: $  17.2\% $ \\  $i$: $  13.9\% $ \\  $z$: $  18.9\% $\\  $y$: $  24.8\% $ } & \makecell{$ \mathbf{all:  91.1\%}$ \\  $r$: $  82.8\% $ \\  $i$: $  86.1\% $ \\  $z$: $  81.1\% $\\  $y$: $  75.2\% $ }  \\ \cline{2-4}
\end{tabular}
\label{tab:cm22_dtmin_4}
\end{table}

\begin{table}
\caption{Set C: $\dtt \sim  U[3.0,24.0]~{\rm and}~ \mut \sim U[1/3,3.0]$}
\centering
\begin{tabular}{ccc|c|}
\multicolumn{2}{c}{} & \multicolumn{2}{c}{\textbf{Predictions}} \\ \cline{3-4}
\multicolumn{2}{c|}{} & \textbf{unlensed} & \textbf{lensed} \\  \cline{2-4}
\multirow{4}{*}{\rotatebox[origin=c]{90}{\textbf{Truth}}} &\multicolumn{1}{|c|}{\textbf{unlensed}} & \makecell{$ \mathbf{all:  96.7\%}$ \\  $r$: $  91.5\% $ \\  $i$: $  95.5\% $ \\  $z$: $  92.7\% $\\  $y$: $  84.3\% $ } & \makecell{$ \mathbf{all:  3.3\%}$ \\  $r$: $  8.5\% $ \\  $i$: $  4.5\% $ \\  $z$: $  7.3\% $\\  $y$: $  15.7\% $ }  \\ \cline{2-4}

& \multicolumn{1}{|c|}{\textbf{lensed}} & \makecell{$ \mathbf{all:  10.7\%}$ \\  $r$: $  19.1\% $ \\  $i$: $  15.8\% $ \\  $z$: $  21.0\% $\\  $y$: $  26.6\% $ } & \makecell{$ \mathbf{all:  89.3\%}$ \\  $r$: $  80.9\% $ \\  $i$: $  84.2\% $ \\  $z$: $  79.0\% $\\  $y$: $  73.4\% $ }  \\ \cline{2-4}
\end{tabular}
\label{tab:cm22_dtmin_3}
\end{table}

\subsection{Early detection}
\label{sec:early_detection}
\begin{figure}
 \centering
\includegraphics[width=\linewidth]{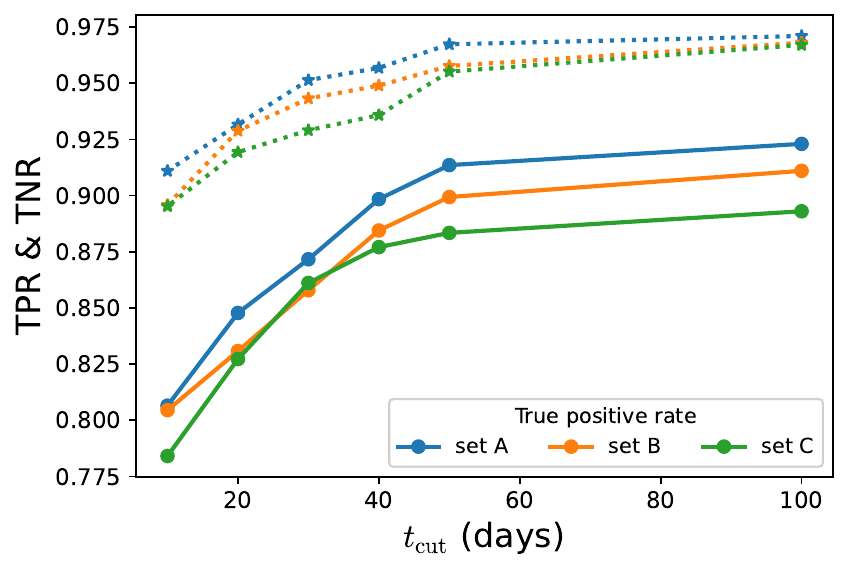}
\caption{Classification results for the light curves truncated at different $t_{\rm cut}$ values. True-positive rates (TPR; solid curves) and true-negative rates (TNR=$1-$FPR; dotted curves) are plotted for four different choices of $t_{\rm cut}$ (note that the far-right points corresponds to the results already described in Section \ref{sec:full_lc} and Tables \ref{tab:cm22_dtmin_5}-\ref{tab:cm22_dtmin_3}). The results for sets A, B, and C are shown with blue, orange, and green colours, respectively. As we truncate the light curves at lower epochs, both TPR and TNR decrease, and therefore FPR increases.
}
\label{fig:class_tcut}
\end{figure}

In the previous section, we show that we could classify the systems well, but we had to use light curves until $100$ days from the first observation in each band. However, our aim is to detect SNe early enough ---preferably in the  rising phase of the light curve--- so that the follow up can resolve them and subsequently measure the time delays. However, the nature of our method limits how early we can detect. In this section, we explore the minimum length of light-curve observations required after the first detection to confidently determine that the SN is lensed. We refer to this duration as $t_{\rm cut}$. (The previous section provides results for $t_{\rm cut}=100$ days.) Therefore, we cut the light curves at an earlier epoch, $t_{\rm cut}$ days from the epoch of first observation. We keep $t_{\rm cut}$ the same for all bands \footnote{Cutting the long processed light curves (as in Figure \ref{fig:lc_inp_b}) at $t_{\rm cut}$ and feeding them to the CNN would provide information as to the peak location, which might actually be beyond $t_{\rm cut}$. Therefore, for a given $t_{\rm cut}$, we always smooth the mag vs time light curves considering data up to the first observation after $t_{\rm cut}$ and then cut the smoothed light curve at $t_{\rm cut}$. Finally, we normalise the flux vs time light curves and feed them to the CNN.} and try to perform the classification separately for different choices of $t_{\rm cut}$, such as at $=50, 40, 30, 20,$ and $10$ days, to check how early we can detect the lensed systems with sufficient accuracy. For our goal of triggering a follow up, we only need the unlensed versus lensed classification, and do not require the $3 \times 3$ classification (unlensed vs double vs quad), which is more difficult.

Instead of the confusion matrix for individual sets, here we present the results in Figure \ref{fig:class_tcut}, which shows true positive (solid curves) and true negative (dotted curves) detection rates as a function of $t_{\rm cut}$. Indeed, the classification accuracy remains reasonably high until $t_{\rm cut} \sim 20$ days, with a TPR of above $82\%,$ while maintaining an FPR of below $9\%$ for the most difficult set C. Therefore, this method shows the potential for detecting the lensed cases early enough to trigger a follow up.

\subsection{Classifying LSST set}

\begin{table}
\caption{Details of the LSST-like systems recovered by models trained on different controlled sets.}
\centering
\begin{tabular}{|c|c|c|c|c|}
\cline{3-5}
\multicolumn{2}{c|}{} & \multicolumn{3}{c|}{Model trained on} \\
\cline{3-5}
\multicolumn{2}{c|}{} & set A & set B &set C \\
\hline
\multirow{3}{*}{Recovery} & all & $ 28.0 \%$ &$ 28.1 \%$  & $ 30.2 \%$\\ \cline{2-5}
 & double & $ 29.5 \%$  &  $ 29.2 \%$ & $ 31.3 \%$ \\ \cline{2-5}
  & quad &  $ 21.9 \%$ & $ 23.3 \%$ &  $ 25.9 \%$ \\ \hline \hline

\multirow{3}{*}{\makecell{$\dt_{\rm med}$ of \\ recovered \\ systems}} & all &  $ 4.5 $ &  $ 4.4 $  &  $ 4.1 $  \\ \cline{2-5}
 & double &  $ 4.8 $  &  $ 4.7 $  & $ 4.6 $  \\ \cline{2-5}
  & quad &  $ 2.6 $  &  $ 2.1$ & $ 1.6$ \\ \hline

\end{tabular}
\tablefoot{The top part presents the percentage of mock LSST unresolved systems recovered by models trained on different sets -- A, B, and C. The three subrows show the recovery rate for the entire recovered sample, and then divided into doubles and quads.
The bottom part shows the median time delays of the recovered systems, again divided into three sets -- all, doubles, and quads from top to bottom. For comparison, the input LSST sample has $\dt_{\rm med}=2.0, 2.4,1.1$ days, respectively, for the full sample, doubles, and quads. It is evident that the models can recover the doubles better than the quads since the former systems have typically larger time delays.}
\label{tab:class_set_lsst}
\end{table}

\begin{figure}
 \centering
\includegraphics[width=0.5\textwidth]{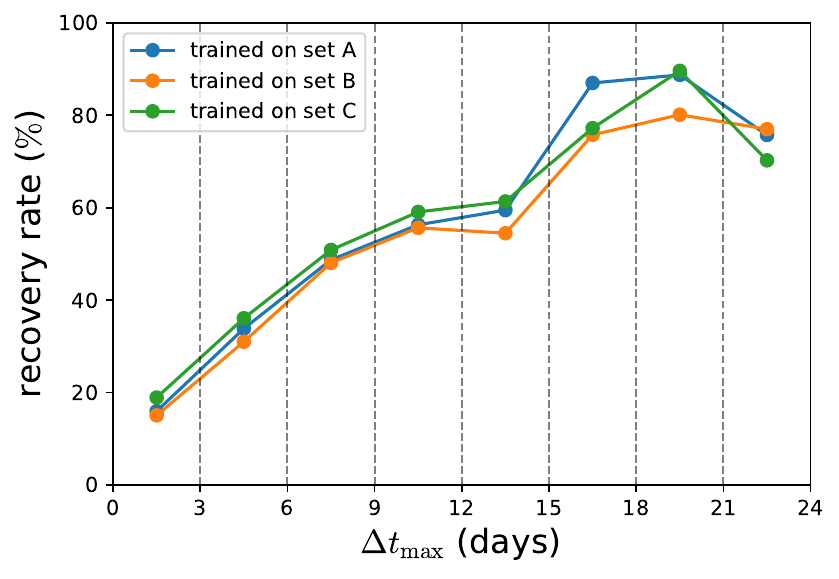}
\caption{Recovery rate for the unresolved systems in set LSST ---which have good multi-band light curves available--- divided into eight time-delay bins. Three colours -- blue, orange, and green -- represent the results obtained using the model trained on the control sets A, B, and C, respectively.}
\label{fig:rec_td_dist}
\end{figure}

Finally, we test the models ---which are trained on the controlled sets A, B, and C--- on an LSST-like sample of unresolved systems (i.e. without any restrictions on $\mu$ or $\dt$). We focus on unlensed vs lensed classifications to avoid any confusion in detection of doubles and quads. The FPRs of these models are already given in Tables \ref{tab:cm22_dtmin_5}--\ref{tab:cm22_dtmin_3} (top right element); for example, the FPR for these models is $\sim 3\%$ when  data from all four bands are used. This exercise unambiguously determines the fraction of unresolved LSNe Ia in LSST that these models can detect.

For simplicity, here we limit our analysis to light curves observed over a period of 100 days, specifically focusing on systems that have been well-monitored across all four LSST bands --- $r$, $i$, $z,$ and $y$. The top part of Table \ref{tab:class_set_lsst} summarises what fractions of the unresolved systems are recovered by the model trained on sets A, B, and C (i.e. TPR as all the test samples are lensed). The first three rows present the TPR considering all systems, only doubles, and only quads, respectively. For all three sets, roughly $\sim 30\%$ of all the systems are recovered. However, the recovery rate slightly improves left to right (from set A to C) when the training set includes lower time-delay systems, which is expected because most of the LSST samples have $\dt \lesssim 2$ days.\footnote{When testing with models trained on a few other sets, where lower $\dt$ and/or $\mu$ systems are included, we found minimal improvements in the recovery rate but at the expense of greater FPR. Therefore, we do not report the results for these additional sets.} Training on set C ($\dt_{\rm min}=3$ days) produces the best recovery rate: ~$31.0\%$. Interestingly, for all three sets, doubles are recovered at a better rate than the quads, because the former typically have larger time delays.

The bottom part presents the median time delay ($\dt_{\rm med}$) of the systems that are recovered by the models trained on different sets. Again the three rows here consider all systems, doubles, and quads separately. For comparison, the input LSST sample has $\dt_{\rm med}=2.03, ~2.37,~1.10$ days for all systems, doubles, and quads, respectively. It is clear that the recovered systems have higher time delays as compared to the total input sample.

Figure \ref{fig:rec_td_dist} explicitly shows the recovery rate of the systems across different $\dt$ bins. The results for models trained on the three controlled sets, shown in three different colours, exhibit a generally consistent pattern with each other. However, the model trained on set C demonstrates slightly superior performance on low time-delay systems. As anticipated from Table \ref{tab:class_set_lsst}, the recovery rate overall improves for systems with higher time delays. For example, for systems with $\dt <3$ days, the recovery rate ranges between $16$ and $20 \%$, whereas it exceeds $\sim 60 \%$ for systems with $\dt$ between $9$ and $12$ days. Moreover, approximately $80-90\%$ of the systems with $\dt >15$ days could be recovered.
We note that the recovery rate for the LSST set, even for higher time-delay systems, is typically slightly lower than in the control sets. This difference arises because LSST samples include systems with $\mu$ smaller than the lower limit on the control sets. These systems are generally more challenging to detect solely based on the shape of the light curves.

\subsection{Time-delay estimation}

\begin{figure*}
 \centering
 \subfigure[set A]{
\includegraphics[width=0.32\linewidth]{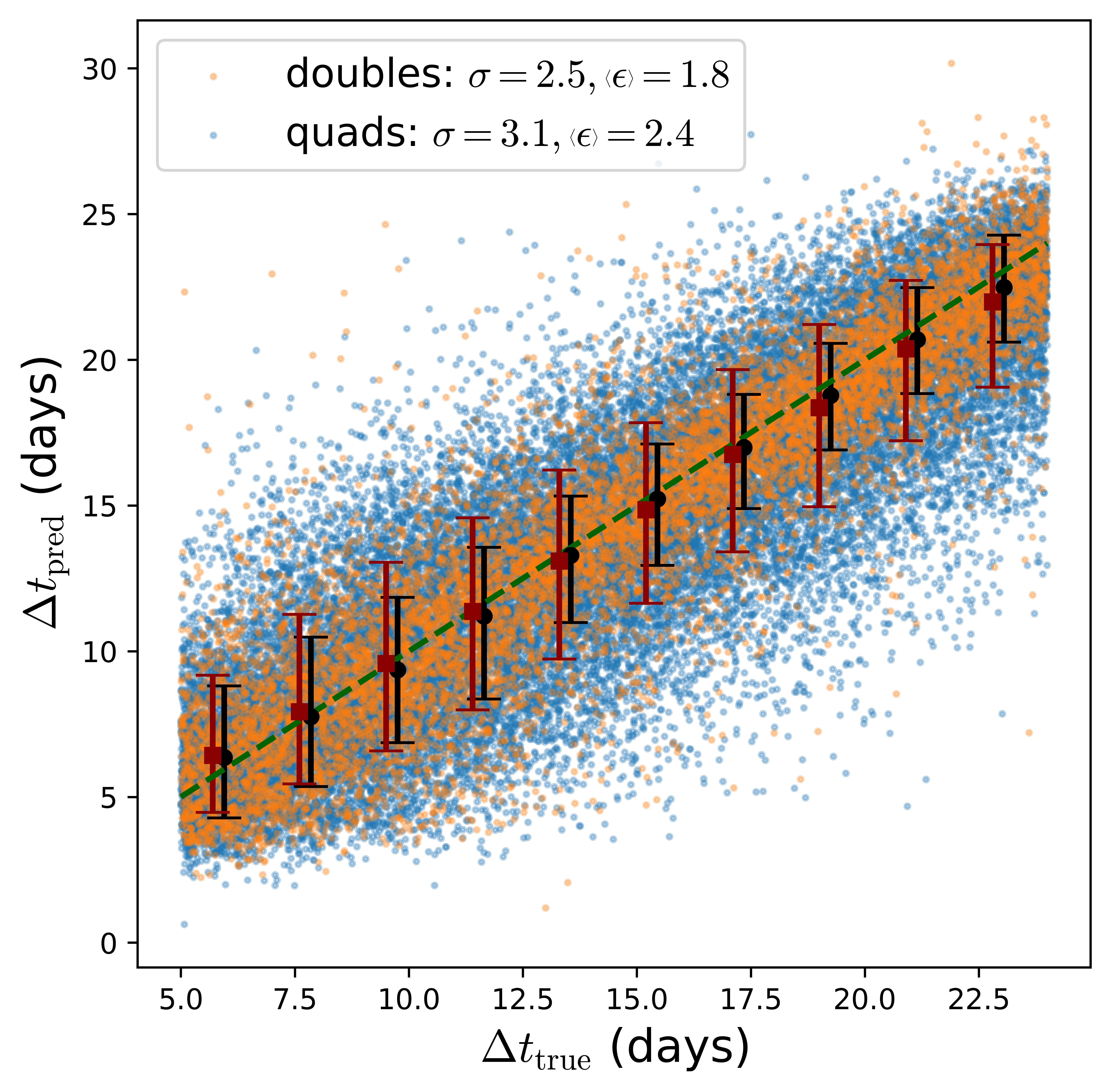}}
 \subfigure[set B]{
\includegraphics[width=0.32\linewidth]{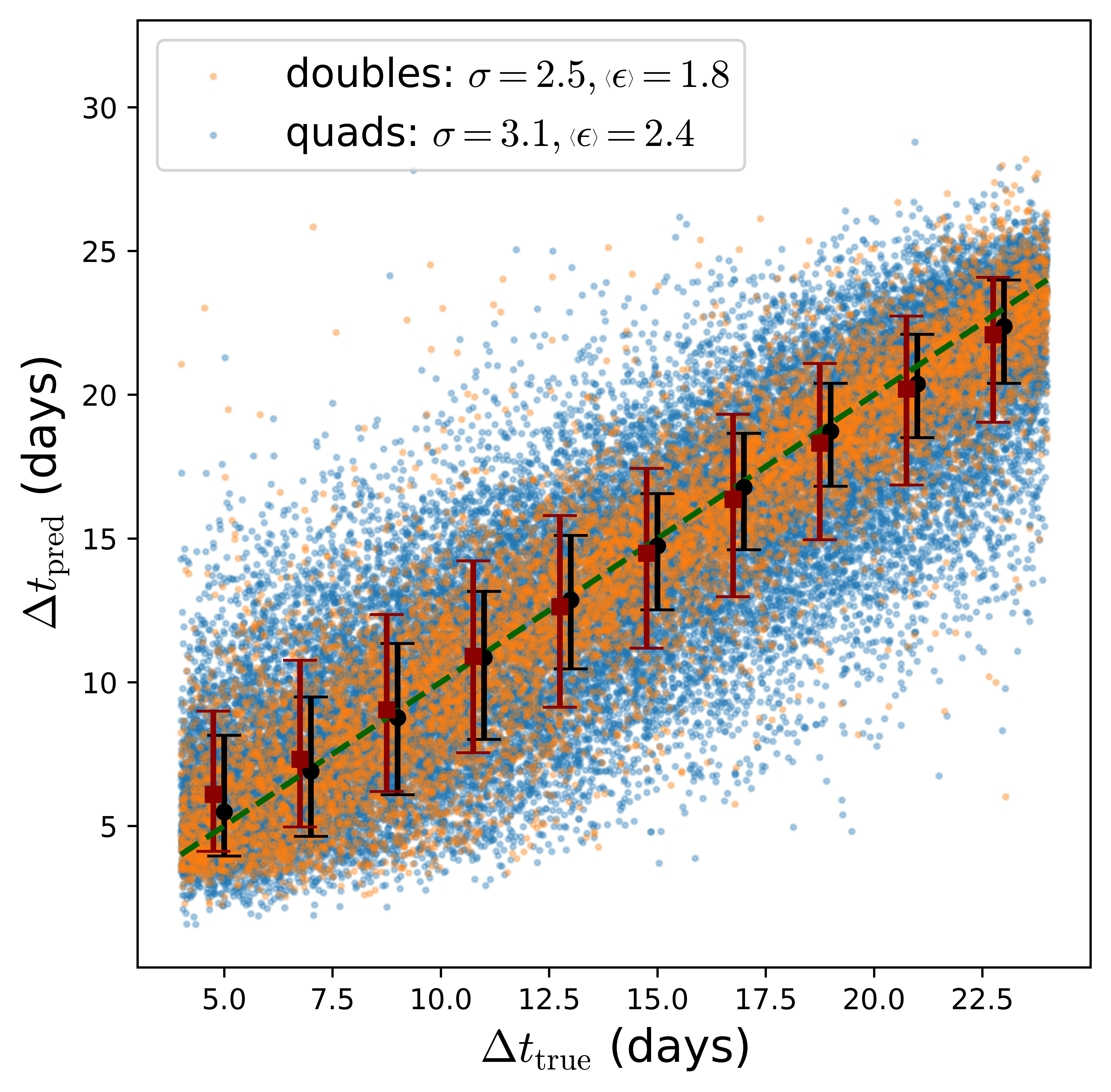}}
 \subfigure[set C]{
\includegraphics[width=0.32\linewidth]{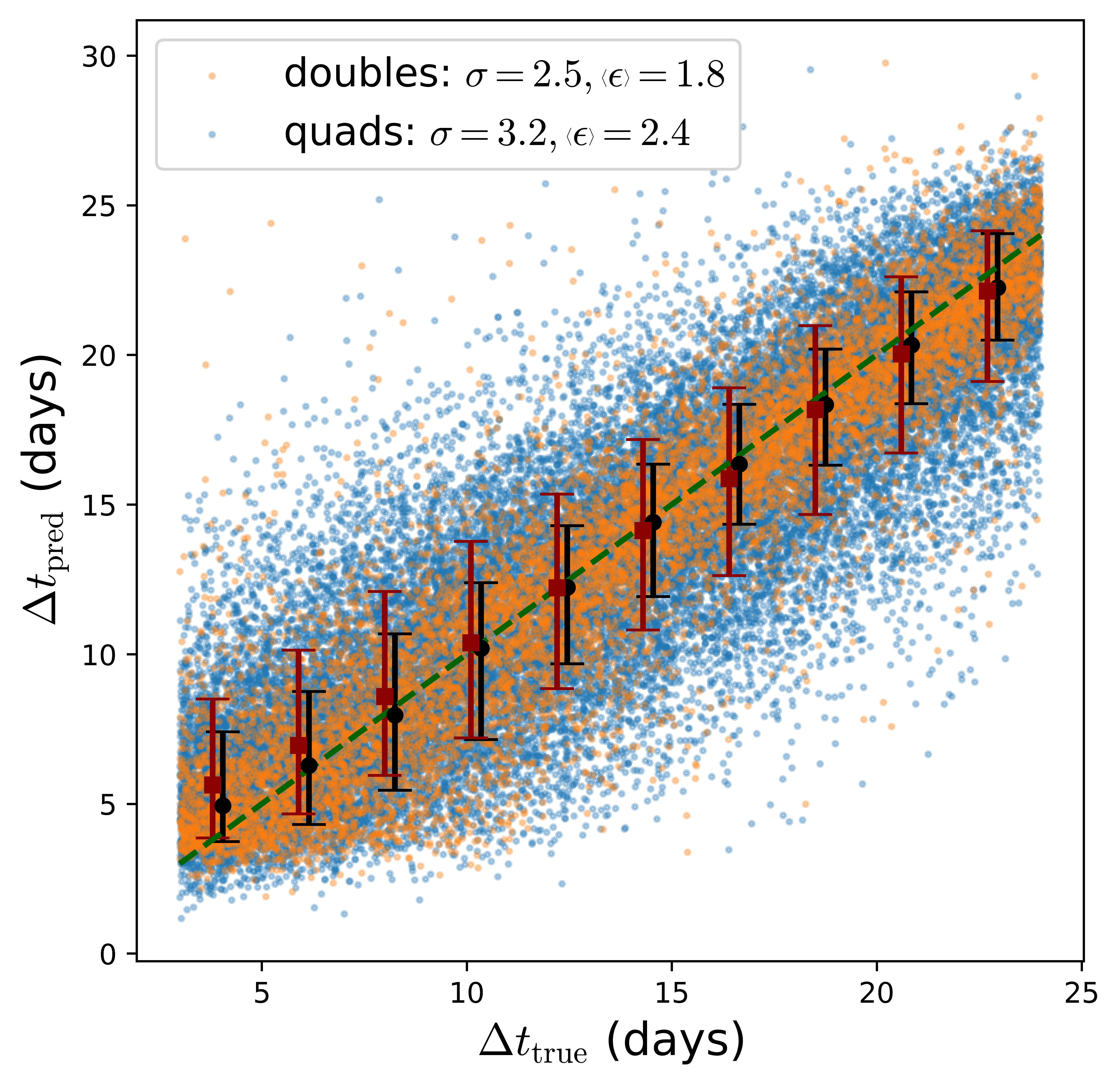}}
\caption{Time-delay estimates from blended joint light curves compared with truths for sets A, B, and C. All predictions come from the model trained on set D ---which covers a larger $\dt$ range ($[0.5,30]$ days)--- in order to minimise apparent bias at $\dt$ boundaries. Orange and blue dots in the plots represent individual predictions for double and quad systems, respectively. The standard deviation of residuals ($\sigma$) and the mean of their absolute values ($\left \langle \epsilon \right\rangle$) are quoted in each plot separately for doubles and quads. Samples are split into ten bins, and the $68\%$ CL around the median is shown by error bars, again separately for doubles (black) and quads (red); the two kinds of error bars are plotted with a slight offset along the x-axis for better visualisation. It is evident that time-delay predictions are more accurate for doubles compared to quads, and, as expected, for higher time delays.}
\label{fig:td_est_set_abc}
\end{figure*}

\begin{figure}
 \centering
\includegraphics[width=0.5\textwidth]{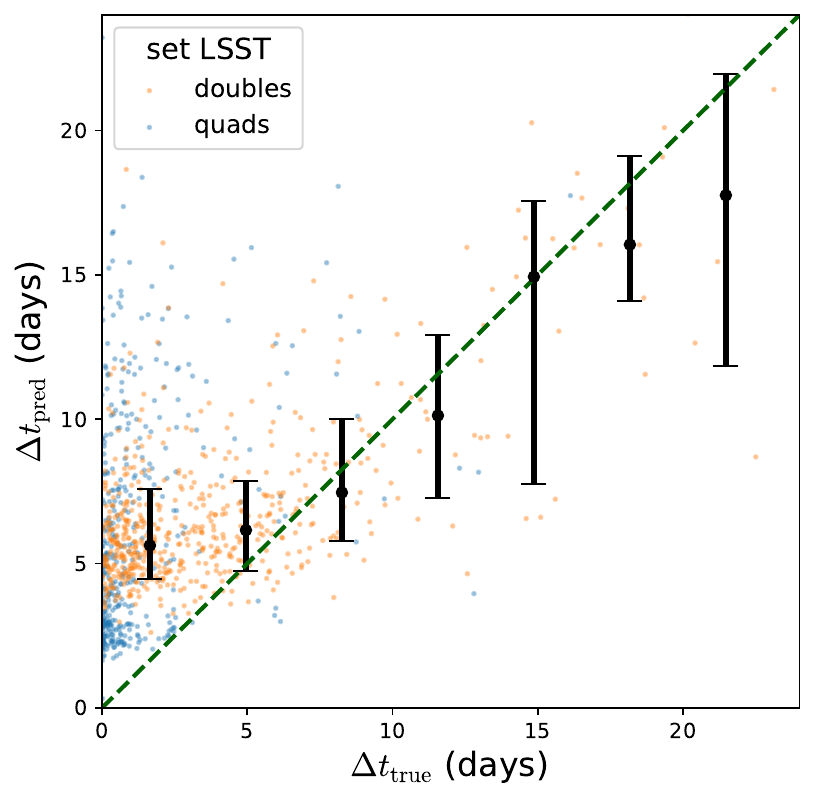}
\caption{Time-delay estimations for the set LSST vs ground truths. Statistics of $\dt_{\text{pred}}$ for the quads (depicted by blue dots) at different bins are omitted, as the $\dt$ values for quads are typically very low, and we lack sufficient numbers at higher $\dt$. Therefore, we present the median and $68\%$ confidence level of $\dt_{\text{pred}}$ at different $\dt_{\text{true}}$ bins only for the doubles. }
\label{fig:td_est_set_lsst}
\end{figure}

Ideally, our goal is to detect the lensed systems using the light curves early enough (not too long after the the peak of the first image) so that follow-up observations can resolve the multiple images (and thus determine if they are doubles or quads) and measure the time delays. However, as we need long light-curve measurements to confidently detect the lensed cases, the time-delay measurements might not be possible for many systems. In such cases, the ability to measure time delays directly from the light curves becomes valuable, and that is the focus of this section.

We train separate networks for predicting time delays in double and quad samples. This assumes proper classification of the system (as a double or quad), either through follow-up observations or a near-perfect classification process, before directing it to the corresponding model. As mentioned above, both networks share the same architecture until the last layer, which has three nodes for quads and one node for doubles, employing a linear activation (or no activation) function.

Figure \ref{fig:td_est_set_abc} compares the estimated time delays against the true values for the three sets --- A, B, and C --- placed in three panels. In each panel, individual $\Delta t$ estimations are represented by small dots, with orange and blue dots corresponding to time delays in doubles and quads, respectively. The green dashed line signifies $\Delta t_{\rm pred}=\Delta t_{\rm true}$, and it is evident that the predicted time delays roughly align with this. This indicates the feasibility of estimating time delays solely from joint light curves, even with LSST-like cadence and noise.

We further segment the time delays into ten equally spaced bins, separately for doubles and quads. For each bin, we calculate the median of the predicted time delays and the $68\%$ confidence level around it, depicted by error bars in green for doubles and red for quads. As anticipated, predictions are more accurate for higher time delays.

It is worth noting that, even after training on a set with a broader $\Delta t$ range (set D), a slight over-prediction is observed at the lower end of $\dt$, which is particularly visible in the right panel of Figure \ref{fig:td_est_set_abc}. This is somewhat inevitable as, when approaching very small time delays, predictions become biased due to the inability to accommodate negative $\dt$.

Importantly, predictions are notably more accurate for double systems compared to quads. Double systems, having only one time delay, exhibit better predictions, while quads, with three time delays, pose a greater challenge to the network. Unlike \cite{misha2022}, we do not impose a minimum time-delay separation between successive images, guided by the distribution of time delays in unresolved systems. This choice makes time-delay estimations for quads considerably more challenging than for doubles.

In each panel, we also provide the standard deviation ($\sigma$) of the residuals and the mean of their absolute values ($\left \langle \epsilon \right\rangle$), separately for the double and quad samples. Across all three sets, we observe that $\sigma$ falls within the range of $3.1$ to $3.2$ days, and $\left \langle \epsilon \right\rangle$ is approximately $2.4$ days for quads. For doubles, $\sigma$ is around $2.5$ days, with $\left \langle \epsilon \right\rangle$ at approximately $1.8$ days. In summary, based on the controlled sets, we show in this section that indeed time-delay estimation is possible from only the joint light curves for the controlled sets.

In our final assessment, we test the model on the LSST set, where the majority of systems exhibit low values of $\Delta t$ and $\mu$ (the median values of $\Delta t_{\text{max}}$ and corresponding $\mu$ are $2.0$ days and $0.4,$ respectively). The predicted time delays are plotted against the ground truths in Figure \ref{fig:td_est_set_lsst}. The estimated time delays for the quads (blue dots) deviate substantially from the true values, reflecting the challenge in accurate prediction, since quads often entail very small time delays (with a median of $\Delta t_{\text{max}}\approx 1.1$ day). Consequently, we abstain from presenting statistics across different $\Delta t$ bins for the quads. On the other hand, the predictions for doubles (orange dots) fare somewhat better. Nonetheless, the model exhibits a tendency to overpredict for low time-delay systems, as expected from the above analysis. Even for moderate time-delay systems, a notable bias is observed, with more values underpredicted than otherwise. This bias is visually evident from the error bars, which depict the median and $68\%$ confidence level around it across different $\Delta t$ bins. Notably, the error bars for the LSST dataset are somewhat larger than those for the controlled sets shown in Figure \ref{fig:td_est_set_abc}. This discrepancy can be attributed to the distribution of $\mu$ in the LSST dataset, which significantly falls below the lower limit of training set D.
Overall, this evaluation underscores the inherent difficulty in estimating time delays solely from the shape of joint light curves of unresolved systems in LSST, as these systems often exhibit low time delays and magnification ratios.

\section{Summary and discussion}
\label{sec:conclusion}
Gravitationally lensed supernovae {}(LSNe) hold significant promise as vital cosmic probes in the near future. The primary challenge of their rarity can be overcome by identifying the abundant unresolved LSNe anticipated in wide-field surveys. In this article, we show how we can leverage the lensing features in the observed light curves to discern between unresolved LSNe Ia and their unlensed counterparts in LSST using a straightforward 1D CNN approach.  Joint light curves of unresolved SNe Ia comprise the positive examples, while the negative examples are represented by the intrinsic light curves of (unlensed) SNe Ia. To simulate LSST-like mock light curves, we employ the cadence distribution and 5$\sigma$ depth from the \texttt{baseline v3.2} observing strategy. Details of the simulation process are provided in Section \ref{sec:train_test_sim}. Out of six LSST filters, we consider four, namely $r$, $i$, $z,$ and $y$, because of their better cadences.

Based on OM10 simulations, we show that LSST will observe approximately three times as many unresolved LSNe Ia as resolved ones. However, for nearly half of these systems, the maximum time delay ($\dt$) is expected to be less than $2$ days. It is crucial to avoid the limit as $\dt$ approaches zero or extreme magnification ratios ($\mu$) in the training process, as in both scenarios, the distinctive lensing signature within the joint light curves diminishes, rendering it nearly impossible for any method to distinguish lensed cases based solely on light-curve shapes. To ensure robustness in our training process, we exclude instances with low values of $\dt$ and $\mu$. Initially, we focus on controlled datasets, denoted A, B, and C. These datasets adhere to specific criteria: both $\dt$ and $\mu$ are uniformly sampled, with $\mu$ following a uniform distribution in the range $[1/3,3]$ in all sets, and the minimum allowed $\dt$ is 5, 4, and 3 days, respectively, for sets A, B, and C. The outcomes for these controlled sets are provided below.
\begin{itemize}
    \item We observe that the classification between unlensed, double, and quad can maintain a well-controlled false-positive rate (FPR). However, there is notable confusion while distinguishing some quads from doubles. This confusion arises due to the fact that joint light curves of certain quads may closely resemble those of doubles, particularly when the separations between time delays are not significant enough.

    \item Hence, the classification between unlensed and lensed cases appears more meaningful. For systems with `well-observed' light curves across all four bands, we achieve a recovery rate of $\sim 90\%$ with a FPR of $\sim 3\%$ for all three sets. Even when considering only a single-band light curve observed among the filters $r$, $i,$ and $z$, the true-positive rate (TPR) remains above $80\%$, with the FPR staying at $\sim 4-7\%$. Consequently, the confident detection of unresolved lensed systems is feasible even with just a single-band light curve.

    \item Early detection plays a pivotal role in initiating follow-up observations. Our demonstration illustrates that even with light curves truncated at 20 days from the initial observation epoch, the reliable detection of lensed cases is feasible with FPR $\lesssim 8\%$ for all three sets. This underscores ample opportunity for triggering follow-up observations aimed at first confirming and then precisely measuring the time delays.

    \item In scenarios where follow-up observations are not possible, we can still estimate the time delays directly from the joint light curves provided the systems are accurately classified as doubles or quads. Time-delay estimations are notably more precise for doubles, as they possess a unique time delay. \footnote{Note that all the results presented above, listed as separate points, are confined to the controlled sets where we exclude low $\dt$ and $\mu$ systems from training and testing; see Table \ref{tab:set_abc}.}
\end{itemize}

Finally, after training these models on controlled sets (A, B, and C), we test their performance on the full LSST sample (which has no restriction on $\dt$ or $\mu$). We manage to recover nearly $30\%$ of these systems provided their light curves are well observed in the four bands. Based on the testing conducted on the controlled sets, we are already aware that the FPR would remain around $3\%$ for these cases. As expected, the recovery rate is higher for larger time-delay systems. For instance, we recover approximately $60\%$ and $80-90\%$ of the systems with time delays ranging between 9 and 12 days, and greater than 15 days, respectively.
Subsequently, we apply the model trained on set D to the set LSST for time-delay estimation. The result reveals challenges in achieving accurate time-delay measurements, characterised by biases and significant uncertainties, even for doubles. This difficulty stems from the predominance of low $\Delta t$ and $\mu$ values in the LSST unresolved samples. Consequently, relying solely on joint light curve data for time-delay estimations may not yield reliable results.

 Note that our method is specifically applicable to SN Ia light curves for distinguishing between lensed and unlensed SNe Ia. Therefore, this approach relies on the accurate identification of SN Ia events beforehand. Current algorithms, such as those of \citet{Lochner2016,Alves22}, demonstrate a high level of precision (as high as $\sim 90\%$, depending on the observing scenario) in detecting SNe Ia among other transients based on photometry. However, some contamination remains, and our method should be adapted to the other types of transients. Additionally, we need to ensure that unresolved LSNe Ia are detected by these algorithms, which have been exclusively developed for unlensed SNe Ia. Given that unresolved LSNe Ia are expected to exhibit short time delays (most with $\lesssim 2$ days), it is likely that the majority of them will be identified.
Addressing these challenges will require rigorous studies beyond the scope of the current work, and is left to future works. These issues are not only relevant to our method but also to other approaches, such as the magnification method, for detecting unresolved SNe Ia in large time-domain surveys like LSST. Therefore, this future work has broader implications for methods relying on photometry alone.

Overall, our approach of detecting the unresolved LSNe Ia solely based on the shape of the blended light curves shows great potential for the upcoming LSST survey. Consequently, the recovery rate (of all unresolved systems) is fundamentally limited by two factors. Firstly, we need well sampled light curves. Therefore, we follow the set of `good observation' criteria, as described in Section \ref{sec:lsst_cadence_noise}, to select the well-monitored systems and leave out those that could not be sufficiently monitored, for example because of season gaps. As these criteria are mostly ad hoc, we have the flexibility to relax them, allowing the inclusion of a larger fraction of unresolved systems. This adjustment will enable us to explore the limit on these criteria, without significantly sacrificing the accuracy of the results.

Secondly, as expected, this method struggles to detect systems with low time delays, which is significant considering that most unresolved systems in LSST will likely have low time delays. Fortunately, lower time-delay systems often exhibit higher total magnification, where the traditional magnification method (and possibly the colour--magnitude diagram too) is expected to perform better. On the contrary, this method performs well for systems with time delays exceeding 5-6 days. Therefore, this approach complements the magnification method.

It is important to note that although our FPR can be as low as $\lesssim 3\%$ under optimal conditions, it can still significantly affect the precision (the ratio of true positives among all positive outcomes) due to the inherently low lensing probability in reality. For example, for a lensing probability of $10^{-4}$, an FPR of \( 3\% \), together with our TPR of $\sim 30\%$ for the LSST dataset, we obtain a precision of  $\sim 1/1000,$ giving rise to  about $ 1000$ false alarms for every true detection. Thus, achieving a lower FPR is crucial. A reduction to an FPR of $\sim 0.1\%$ under the same conditions would significantly decrease the false alarm rate to a more manageable value of  $30$.

 Another promising way to significantly improve the classification efficiency is by incorporating imaging and/or spectroscopic data alongside the light curves. For instance, in lensed cases, we anticipate the presence of the lensing galaxy near the SN. Moreover, in many instances, the host galaxy might be evident in the reference image. Additionally, the shape of an unresolved SN will not necessarily appear point-like; rather, it could be elongated in the difference imaging, possibly with its centroid shifting over time as different images dominate the observed light. Furthermore, for type Ia SNe, estimating the magnification is feasible as they can serve as standardisable candles. This allows us to verify whether the observed brightness aligns well with the photo-$z$ estimation of the lens. Incorporating this supplementary information should lead to a better classification efficiency with lower FPR.
Another potential path for improvement lies in the network architecture. Instead of employing a CNN, which necessitates fixed-length input data, we can use a long short-term memory (LSTM) network. LSTM networks have the advantage of being able to handle variable-length input sequences and can also accommodate flux uncertainties directly as inputs. These aspects will be further explored in subsequent research endeavors.

It is worth noting that unresolved SNe can be observed up to higher luminosity distances, presenting an opportunity to study the properties of these phenomena at high redshifts. For instance, they can be used to investigate whether or not there is any evolution in the light curves of SNe Ia.

In summary, this study showcases the viability of identifying unresolved LSNe Ia through light-curve observations with LSST using deep-learning techniques. Additionally, we illustrate the potential for extracting time delays solely from these unresolved light curves, enabling the prompt utilisation of such systems delivered by LSST, which may be overlooked by most other detection methods. Although the achieved FPR is below $3\%$ in the optimal scenario, further exploration with more complex network architectures and inclusion of additional datasets may aid in further reducing this rate.

\section*{Data availability}
 The data products (including the catalog of unresolved LSNe Ia, microlensed light curves) are available at Zenedo \href{https://zenodo.org/records/13644602}{DOI 10.5281/zenodo.13644602}. The relevant codes (including {\tt ipython} notebooks) for analyzing the unresolved systems and constructing the training and test sets comprising their blended joint light curves are available at \href{https://github.com/deltasata/Unresolved_LSNeIa_in_LSST}{https://github.com/deltasata/Unresolved\_LSNeIa\_in\_LSST}.

\begin{acknowledgements} 
We thank Masamune Oguri for providing the updated OM10 simulation code and for offering valuable comments. We also thank Taeho Ryu and Kaitlyn Szekerczes for their assistances in setting up the code. SB acknowledges the insightful discussions with Georgios Vernardos, Mikhail Denissenya and Wuhyun Sohn on various aspects of this project. SB acknowledges the funding provided by the Alexander von Humboldt Foundation. SHS thanks the Max Planck Society for support through the Max Planck Fellowship. This project has received funding from the European Research Council (ERC) under the European Union’s Horizon 2020 research and innovation programme (grant agreement No 771776). This research is supported in part by the Excellence Cluster ORIGINS which is funded by the Deutsche Forschungsgemeinschaft (DFG, German Research Foundation) under Germany’s Excellence Strategy -- EXC-2094 -- 390783311. SS has received funding from the European Union’s Horizon 2022 research and innovation programme under the Marie Skłodowska-Curie grant agreement
No 101105167 - FASTIDIoUS. AS would like to acknowledge the support by National Research Foundation of Korea NRF-2021M3F7A1082056, and the support of the Korea Institute for Advanced Study (KIAS) grant funded by the government of Korea.
\end{acknowledgements}

\bibliographystyle{aa}
\bibliography{ref1}



\appendix

\section{Source and lens redshift of the unresolved systems in LSST}
\label{app:unres_stat_extra}

\begin{figure*}
 \centering
\includegraphics[width=0.485\textwidth]{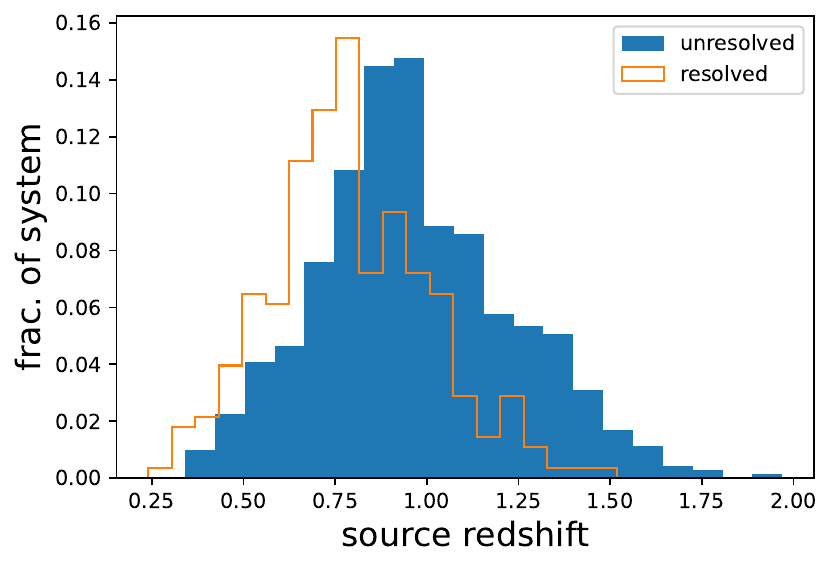}
\includegraphics[width=0.485\textwidth]{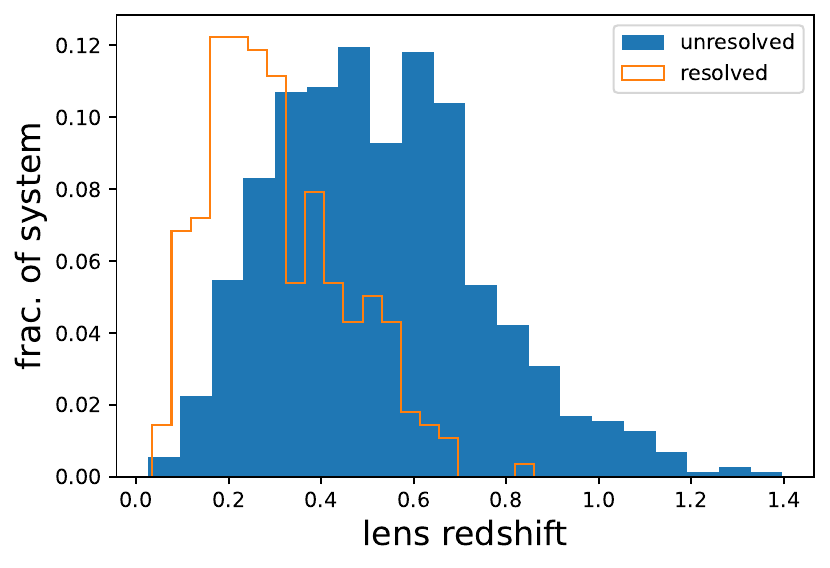}
\caption{The unresolved and resolved LSNe Ia in LSST have been compared in terms of their distributions of source redshift (left panel) and lens redshift (right panel). Since unresolved systems can be observed up to larger luminosity distances, they can accommodate larger source redshifts as compared to the resolved ones. Conversely, their smaller angular separations necessitate larger lens redshifts compared to the resolved systems. }
\label{fig:z_sl}
\end{figure*}

\begin{figure*}
 \centering
\includegraphics[width=0.485\textwidth]{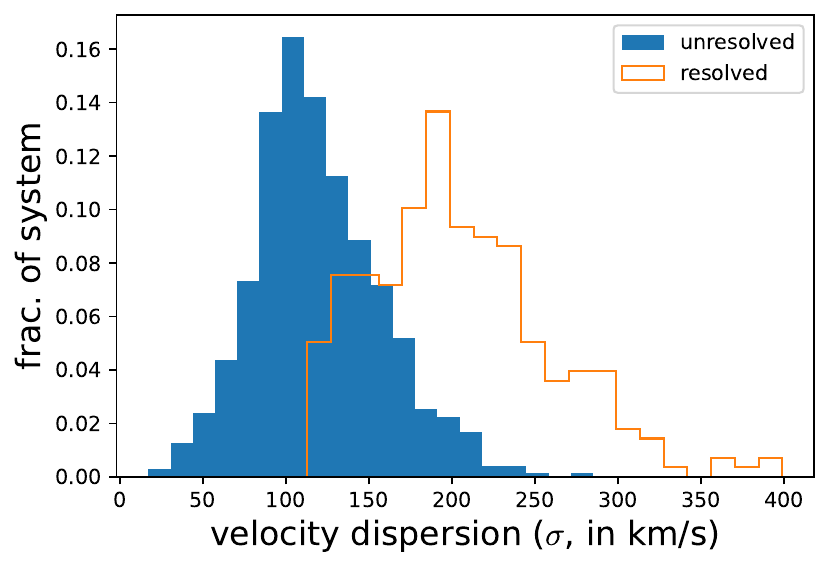}\hspace*{1mm}
\includegraphics[width=0.48\textwidth]{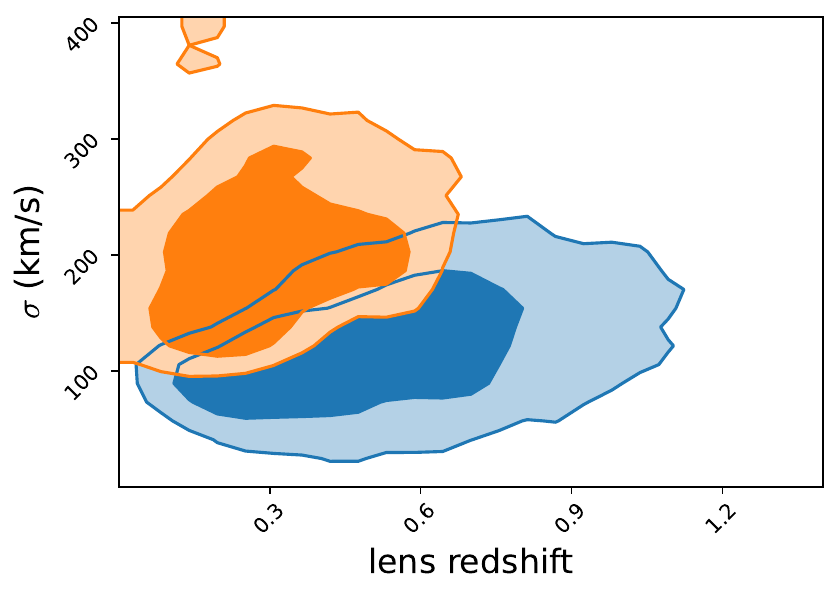}
\caption{The left panel compares the distribution of the velocity dispersion ($\sigma$) in the lens galaxies for unresolved (blue) and resolved (orange) systems. The right panel shows 2D distribution of lens galaxies in the velocity dispersion vs redshift plane, separately for unresolved (blue) and resolved (orange) systems. Darker and lighter shades correspond to the $68\%$ and $95\%$ confidence intervals, respectively. Since velocity dispersion serves as a proxy for the mass of elliptical galaxies, the combined insights from both panels indicate that lenses in unresolved systems are either lower-mass galaxies, similarly massive galaxies at higher redshifts, or a mix of both factors.}
\label{fig:vel_disp_comp}
\end{figure*}

In Figure \ref{fig:z_sl} we compare between the resolved and unresolved LSNe Ia in LSST in terms of their source and lens redshifts in the left and right panel respectively. Unresolved systems, being observable at greater distances, can accommodate larger source redshifts compared to resolved ones. This presents a unique opportunity to investigate the properties of SNe at high redshifts, such as potential evolution in the light curve of SNe Ia. Understanding such evolution is crucial for cosmology, as Type Ia SNe serve as fundamental tools for studying the universe's expansion. On the other hand, the need for low angular separations in unresolved systems, compared to resolved ones, requires smaller Einstein radii. This can be achieved either by having higher lens redshifts or lower lens masses for the same set of sources. The right panel of Figure \ref{fig:z_sl} explicitly shows that unresolved systems tend to have higher lens redshifts compared to resolved systems.

Figure \ref{fig:vel_disp_comp} provides a detailed comparison of lens galaxies for unresolved and resolved systems. The left panel shows the distribution of velocity dispersion ($\sigma$), which serves as a proxy for the mass of elliptical galaxies. As anticipated, unresolved systems are associated with lower-mass galaxies. The right panel presents the 2D distribution of lens galaxies in the $\sigma$ vs $z$ plane, aiming at distinguishing unresolved (blue) and resolved (orange) systems. Darker and lighter shades represent the $68\%$ and $95\%$ credible regions, respectively. This panel indicates that unresolved systems are likely to be associated with either lower-mass galaxies or similarly massive galaxies at higher redshifts compared to resolved systems, though some overlap exists.

\end{document}